\documentclass[review,hidelinks]{elsarticle}

\usepackage{fullpage}
\usepackage{afterpage}

\usepackage{amsmath}
\usepackage{amssymb}

\usepackage{booktabs}
\usepackage[table,x11names,dvipsnames,table]{xcolor}

\usepackage[acronym]{glossaries}
\input{glossary.aux}

\newcommand{\Xtop}[1]{\ensuremath{E(X_\text{DIS#1})}}

\newcommand{\Etx}{\ensuremath{\text{En}_{\text{TX}}}}
\newcommand{\Erx}{\ensuremath{\text{En}_{\text{RX}}}}
\newcommand{\Etop}[1]{\ensuremath{\text{En}_{\text{DIS#1}}}}
\newcommand{\Emin}{\ensuremath{\text{En}_{\text{MIN}}}}

\newcommand{\Ptx}[1]{\ensuremath{E(S{#1})}}
\newcommand{\Prx}[1]{\ensuremath{E(R{#1})}}
\newcommand{\Ptxmax}[1]{\ensuremath{E(max(S{#1}))}}
\newcommand{\Prxmax}[1]{\ensuremath{E(max(R{#1}))}}

\newcommand{\frequency}{\ensuremath{f}}

\newcommand{\Constant}[1]{\emph{Constant}}

\usepackage{tikz}

\usepackage{hyperref}

\journal{Computer Communications}

\usepackage[pdftex]{}
\graphicspath{{images/}{jpeg/}}
\DeclareGraphicsExtensions{.pdf,.jpeg,.png}

\usepackage{paralist}
\usepackage{mathtools}
\usepackage{algorithmic}

\usepackage{caption}
\usepackage{subcaption}

\usepackage[linesnumbered]{algorithm2e}

\begin{document}

\begin{frontmatter}

\title{The Impact of Dual Prediction Schemes on the Reduction \\ of the Number 
of Transmissions in Sensor Networks}

% % \author{\IEEEauthorblockN{Gabriel Martins Dias, Boris Bellalta and Simon Oechsner}
% % \IEEEauthorblockA{~\\Department of Information and Communication Technologies\\ 
% % Universitat Pompeu Fabra, Barcelona, Spain\\
% % Email: \{gabriel.martins, boris.bellalta, simon.oechsner\}@upf.edu}
% % }
% % % make the title area
% % \maketitle

% \author{Gabriel Martins Dias\corref{mycorrespondingauthor}}
% \cortext[mycorrespondingauthor]{Corresponding author 
% \ead{gabriel.martins@upf.edu}}
% \address{Pompeu Fabra University, Barcelona, Spain}
% 
% \author{Boris Bellalta}

\author{Gabriel Martins Dias\corref{mycorrespondingauthor}}
\cortext[mycorrespondingauthor]{Corresponding author}
\ead{gabriel.martins@upf.edu}

\author{Boris Bellalta\corref{}}
\author{Simon Oechsner\corref{}}
\address{Pompeu Fabra University, Barcelona, Spain}

\begin{abstract}
%\boldmath

Future \gls{IoT} applications will require that billions of wireless devices 
transmit data to the cloud frequently. 
However, the wireless medium access is pointed as a problem for the next 
generations of wireless networks; hence, the number of data transmissions in 
\glspl{WSN} can quickly become a bottleneck, disrupting the exponential growth 
in the number of interconnected devices, sensors, and amount of produced data.
Therefore, keeping a low number of data transmissions is critical to incorporate 
new sensor nodes and measure a great variety of parameters in future generations 
of \glspl{WSN}.
Thanks to the high accuracy and low complexity of state-of-the-art forecasting 
algorithms, \glspl{DPS} are potential candidates to optimize the data 
transmissions in \glspl{WSN} at the finest level because they facilitate for 
sensor nodes to avoid unnecessary transmissions without affecting the quality of 
their measurements.
In this work, we present a sensor network model that uses statistical theorems 
to describe the expected impact of \glspl{DPS} and data aggregation in 
\glspl{WSN}.
We aim to provide a foundation for future works by characterizing the 
theoretical gains of processing data in sensors and conditioning its 
transmission to the predictions' accuracy.
% and design a mechanism for data 
% reduction. 
% Then, based on a data study, we show how effective can be its use to reduce the 
% number of transmissions in WSNs and increase their lifetime.
Our simulation results show that the number of transmissions can be reduced by 
almost $98\%$ in the sensor nodes with the highest workload.
We also detail the impact of predicting and aggregating transmissions according 
to the parameters that can be observed in common scenarios, such as sensor 
nodes' transmission ranges, the correlation between measurements of different 
sensors, and the period between two consecutive measurements in a sensor.
\end{abstract}
\glsresetall{}


\end{frontmatter}

\section{Introduction}

Wireless sensor nodes (sensor nodes, for brevity) are small computer devices 
with a radio antenna. 
They are often equipped with sensors that are capable of sensing one or more 
environmental parameters.
As an example, temperature and relative humidity sensors are some of the 
cheapest and smallest sensor chips available and are commonly used in real 
world applications.

Sensor nodes are usually organized as \glspl{WSN} that, at the application 
layer, have two fundamental roles:
\glspl{GW} and ordinary sensor nodes.
Ordinary sensor nodes are typically close to the data origin and may just 
perform default sensing tasks and transmit their measurements via radio to a 
\gls{GW}.
\glspl{GW} are responsible for forwarding the gathered data
to \glspl{WSN}' managers and for disseminating occasional instructions and 
updates to sensor nodes.

The \glspl{WSN}' growth can be described by the increasing number of 
wireless sensor nodes measuring and reporting data to \glspl{GW}, and by the 
diversity of data types transmitted in \glspl{WSN}.
As a consequence of this growth, modern applications of \glspl{WSN} do not 
simply monitor changes in the environment anymore; they also trigger reactions 
to these changes. 
For example, in agriculture, sensed data could be used to apply pesticides after 
detecting that the number of insects exceeded a certain 
threshold~\cite{Jurdak2015}.
Such a threshold, in turn, may vary according to the season or get affected by 
temperature and relative humidity changes during the days.
In these applications, sensor nodes may need a high number of transmissions 
to communicate the number of insects, temperature, and relative humidity.
If the \gls{WSN} cannot handle all transmissions that sensor nodes make, it 
will collapse and end up in significant economic losses.

In other cases, losses can be even worse.
For example, structural health monitoring for aircraft can include engine 
control systems that rely on enhanced data analysis to detect accidents 
and report unexplained phenomena to people responsible for maintaining their 
safety and healthy conditions~\cite{McCorrie2015}.
Extended data analysis usually require, in comparison to the simplest monitoring tasks, 
more parameters and a higher amount of informative data.
Therefore, in these cases, more sensor nodes transmitting higher amounts of 
data might collapse the \gls{WSN} and mispredict accidents, resulting in
living losses.

These situations help us to understand that \glspl{WSN} are data-oriented 
networks, i.e., the data produced by the sensor nodes is their most valuable 
asset~\cite{Stankovic2004}.
However, sensor nodes typically have limited energy resources, and transmitting 
data is the task that drains the most battery power.
Hence, several works addressed sensor nodes' energy management as one of the 
biggest challenges for \gls{WSN} applications~\cite{Borges2014}.
% A potential problem of having a sensor node running out of battery in a WSN is 
% that the exact data they could produce would not to be available anymore. 
% Such an impact can be better understood if we consider that most of the 
% In case a sensor node is responsible for providing a communication bridge 
% between its neighbors, its absence can--in the best case--make the other sensor 
% nodes consume more energy trying to find new routes to the \BaseStation{}.
% However, occasionally, their role will be essential to keep the full 
% connectivity in the WSN.

% In some cases, WSNs are internally divided into clusters, and each cluster has 
% its own Cluster Head (\BaseStation{}) that is responsible for the 
% communication 
% between its sensor nodes and the \BaseStation{}~\cite{Abbasi2007}.

% , \gls{IoT} application 
% scenarios will combine new services and dramatically increase the number of 
% sensing devices, such as, among others, tiny sensors, smartphones, home 
% appliances and vehicles~\cite{Bellavista2013}.
Fortunately, new technologies can harvest energy from solar, mechanical and 
thermal energy sources~\cite{Zheng2011,Amaro2012}, showing that this problem can 
be--at least, partially--overcome in the next years.
However, in addition to the energy consumption of the sensor nodes, the 
efficient use of spectrum resources has been pointed out as one of the key 
challenges that will affect the next generation of wireless networks, for 
instance, WLANs, 4G and 5G 
networks~\cite{Adame2014,Chang2014,Chatzikokolakis2015}.

In this work, we present a \gls{WSN} model for data 
transmissions in monitoring applications.
As future generations of sensor nodes tend to evolve and eventually incorporate 
other \gls{IoT} devices (such as smartphones, home appliances, and vehicles), 
variables as sensor nodes' connectivity and their maximum distance to the 
\gls{GW} are considered in the model.
In the end, focusing on the challenges for the next generations of \glspl{WSN}, 
we use statistical theorems to draw conclusions about how efficient is to adopt 
a \gls{DPS} to reduce the number of data transmissions in a monitoring 
application.
As an outcome, one of the main contributions of this work is a mathematical 
model that formulates the number of transmissions in a \gls{WSN} that 
simultaneously adopts a \gls{DPS} and aggregation to reduce its number of 
transmissions.

Our goal is to provide a sufficiently strong background on which future 
applications can rely to create smarter and more complex systems.
In the next sections, we will evaluate the gains that can be obtained by 
reducing the number of data transmissions and compare with the costs of 
choosing a certain prediction model, considering its expected (in)accuracy and 
the correlation between the measurements made by different sensor nodes.

The rest of the paper is organized as follows:
% Section~\ref{sec:background} details how we model the data using Normal 
% distributions, which is fundamental for understanding the assumptions we 
% will make further;
Section \ref{sec:background} describes how \glspl{DPS} can be adopted in 
\glspl{WSN}, based on previous studies in this field;
Section \ref{sec:model} describes our \gls{WSN} transmission model in 
detail;
Section~\ref{sec:proposed-mechanism} describes how we use our model to 
represent a \gls{DPS} and estimate its benefits in \glspl{WSN};
Section~\ref{sec:experimentation} models the number of transmissions 
and the energy consumption expected when adopting the data aggregation or the 
data prediction methods;
% Section~\ref{sec:use-case} presents the experimental study and the simulation 
% results; 
Section~\ref{sec:related-work} shows the current state of the art, giving a 
wider perspective and highlighting the contributions of this work;
and
Section~\ref{sec:conclusion} shows our conclusions and outlines for future work.

\section{Background - Dual Prediction Schemes}
\label{sec:background}

The term \emph{prediction} can either refer to the process of inferring missing 
values in a dataset based on statistics or empirical probability, or to the 
estimation of future values based on the historical data.
% The latter mechanism is also called \emph{forecast}, and it is the class of 
% predictions we will refer to in this Chapter.
% To predict the future, we must consider a high number of possible outcomes, 
% given the uncertainty about the factors that may impact the scenario under 
% consideration.
% Thus, forecasts differ from other predictions because this range of 
% possibilities tends to be wider than in cases when missing values are inferred.
% 
A \emph{prediction method} $P$ is a deterministic algorithm that produces 
predictions based on two input variables: a set of \emph{observed values} $X$ 
and a \emph{set of parameters} $\theta$.
A \emph{prediction model} $p$ is an instance of a prediction method $P$, 
such that $p_{\theta}(X) = P(X, \theta)$.
Thus, a prediction model can be summarized as $P$ and $\theta$.

The values of $\theta$ can be provided by a utility function that measures 
the predictions' accuracy, models' complexity or information loss.
Thus, \emph{choosing a prediction model} means finding the values of $\theta$ 
that best summarizes the current measurements under the criteria adopted by the 
utility function (e.g., the minimum information loss estimated).
% In Appendix~\ref{chapter:how-choose-prediction-model}, we summarize methods 
% that have been used to choose prediction models in \glspl{WSN}.

To choose a prediction model in a \gls{DPS}, sensor nodes report all the data 
that they have measured to the \glspl{GW} during an \emph{initialization 
phase}~\cite{Santini2006}.
% new prediction models must generated 
% After this phase, prediction models can be chosen. 
The choice of the prediction model may be made independently in \gls{GW} 
and sensor nodes;
alternatively, each sensor node can choose its prediction model and 
transmits the parameters to the \gls{GW};
or, finally, the \gls{GW} may be given the autonomy to choose prediction 
models for all sensor nodes.
The decision about where to choose the prediction model can be taken at 
runtime, and it is not necessary to fix a strategy for the 
whole \gls{WSN}'s lifetime~\cite{Lazaridis2003}.

After choosing (and occasionally sharing) the prediction models, sensor nodes 
can exploit their proximity to the data's origin.
That is, they can compare predicted values with real measurements and transmit 
the actual measurements only if the predictions are inaccurate.
From time to time, if predictions' accuracy gets compromised, sensor nodes and 
\gls{GW} may begin a new \emph{initialization phase} and choose new prediction 
models.

\subsection{Independent model choice}

\begin{figure}[t]
	\centering
	\includegraphics[width=0.8\textwidth]{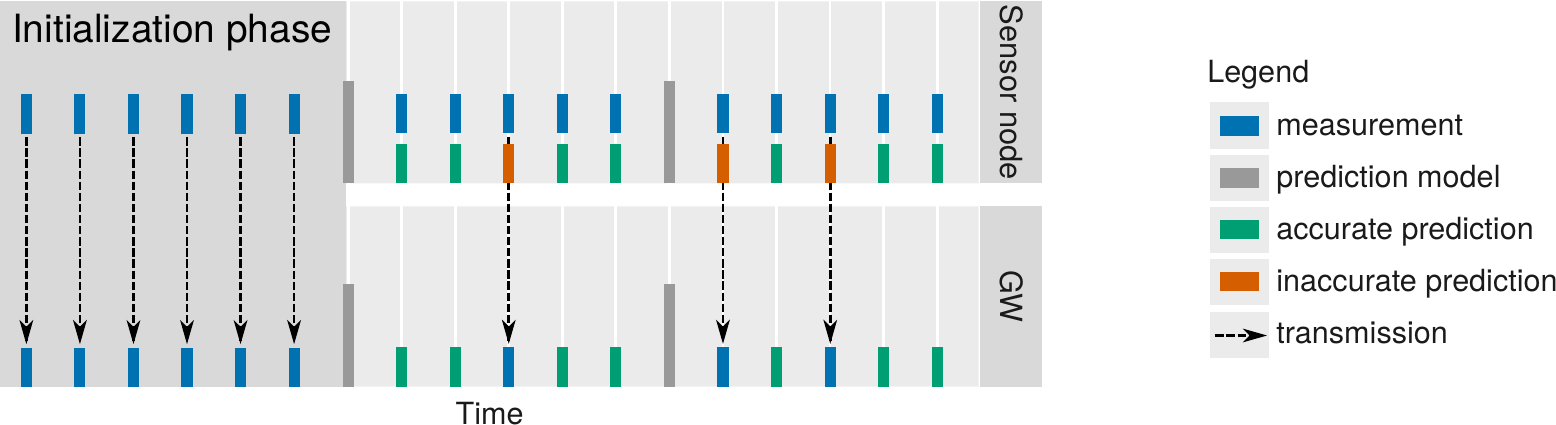}
	\caption{A variant of the \glsentryshort{DPS} with independent model 
generation. The sensor node and the \glsentryshort{GW} can compute the same 
prediction model because they are programmed to use the same data.}
	\label{fig:timeline-dps-independent}
\end{figure}

The \emph{initialization phase} ensures that the \gls{GW} will have complete 
information about the environment before any prediction model is chosen. 
Therefore, after this phase, the \gls{GW} can choose the same prediction 
models 
as sensor nodes without making any new transmission.
Figure~\ref{fig:timeline-dps-independent} illustrates the sensor nodes' and 
\gls{GW}'s behaviors.
New prediction models can be regularly chosen based on the knowledge 
simultaneously available to sensor nodes and \gls{GW}.
As a drawback, the variety of the prediction models is restricted by the memory 
and computing power limitations of sensor nodes.

The \gls{LMS} method has provided accurate predictions in 
simulations where sensor nodes and \gls{GW} generated their prediction 
models independently~\cite{Santini2006,Stojkoska2011,Wu2016}.
For instance, in a particular scenario, only $10\%$ of the measurements would 
be necessary to monitor room temperature accurately~\cite{Santini2006}.

\subsection{Model choice in sensor nodes}

\begin{figure}[t]
	\centering
	\includegraphics[width=0.8\textwidth]{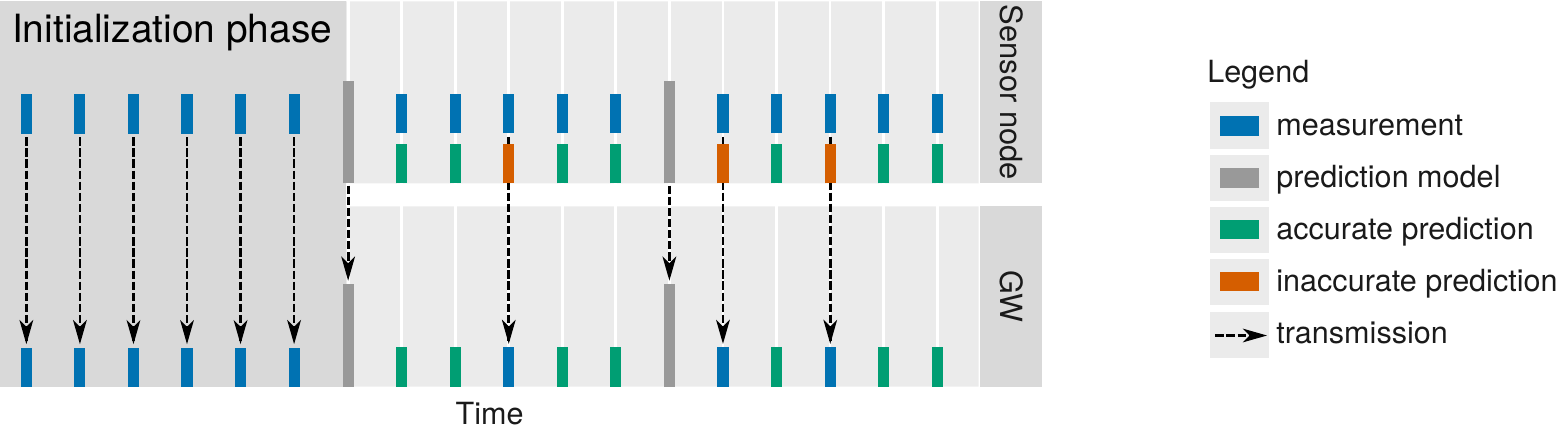}
	\caption{As sensor nodes can overhear their neighbors' data without 
overloading the network or congesting the medium, they may locally decide the 
best prediction method and later inform their decision to the 
\glsentryshortpl{GW}.}
	\label{fig:timeline-dps-sensor-node}
\end{figure}

Alternatively, prediction models can be chosen in sensor nodes, as 
illustrated in Figure~\ref{fig:timeline-dps-sensor-node}.
% (and not in \glspl{GW}, the alternative solution), 
As before, sensor nodes start transmitting all the measurements to 
the \gls{GW}.
However, a new responsibility is assigned to sensor nodes: 
after collecting local measurements and choose a prediction model that fits 
the current environment, they must communicate the prediction model to 
the \gls{GW}.
The main advantage of this mechanism is that sensor nodes can decide for new 
prediction models using all the measured data, instead of using only the 
information that they share with the \gls{GW}.
On the other hand, sensor nodes need extra transmissions to inform 
the \gls{GW} about their decisions.

%%% Automatic ARIMA modeling-based data aggregation scheme in \glspl{WSN}
Simulation results using real data from \glspl{WSN} showed that this approach 
could reduce the number of data transmissions using \gls{AR}, 
\gls{ARIMA}, or \gls{ES} models
% \footnote{In 
% Appendix~\ref{chapter:app-forecasting}, we describe all of these methods in 
% detail.} 
with neither exceeding the constrained memory nor the computational 
resources of typical wireless sensor 
nodes~\cite{LeBorgne2007,Li2013,mccorrie2015predictive}.
%%% Hybrid ARIMA and Neural Network Model for Measurement Estimation in 
%%% Energy-Efficient Wireless Sensor Networks
Alternatively, a hybrid mechanism may improve the quality of the predictions 
if sensor nodes have the autonomy to adopt more complex prediction methods 
(e.g., \glspl{ANN}) when the simplest predictions (e.g., \gls{ARIMA}) 
are inaccurate~\cite{Askari2011}. 
Only in the worst case, if the difference between the measurements and the
predictions using the most complex method also 
exceed the acceptance threshold, sensor nodes are responsible for 
transmitting the real measurements to the \gls{GW}.

More recently, in~\cite{Kho2009,Chen2015}, the authors gave to the sensor 
nodes the ability to take decisions locally using Gaussian Processes and 
Stochastic Gradient Descent regression, which require much higher computational 
power than the traditional methods.
Following the trend of adopting complex prediction methods in sensor 
networks, in~\cite{Cheng2015}, the authors incorporated information theory in 
their analysis and described a method that can accurately predict and evolve 
their prediction models.

\subsection{Model choice in the Gateway}

\begin{figure}[t]
	\centering
	\includegraphics[width=0.8\textwidth]{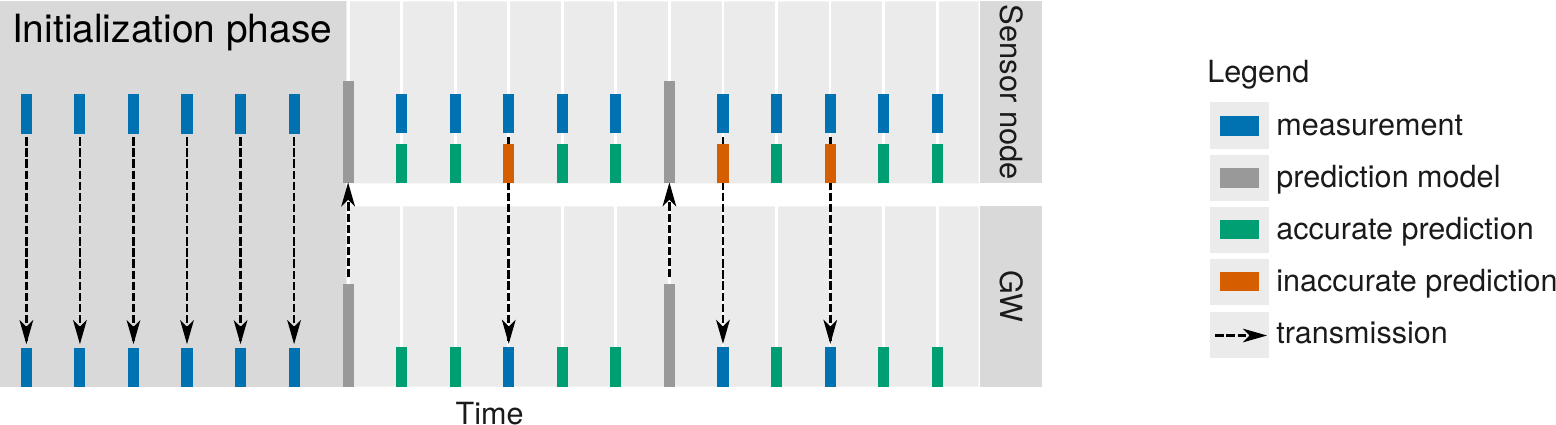}
	\caption{In a \glsentryshort{DPS}, a measurement is only transmitted if its 
forecast is inaccurate. The \glsentryshortpl{GW} may be responsible for 
transmitting new prediction models every time interval after the initialization 
phase.}
	\label{fig:timeline-dps-ch}
\end{figure}

In this type of \gls{DPS}, the \gls{GW} is responsible for periodically 
choosing and transmitting new prediction models' parameters and error 
acceptance 
levels to sensor nodes, as shown in Figure~\ref{fig:timeline-dps-ch}.
Generating the prediction models in the \gls{GW} exploits the asymmetric 
computational power availability in \glspl{WSN}: \glspl{GW} usually have 
cheaper energy sources and more resources (such as memory and processing power) 
than ordinary sensor nodes that mainly measure and report environmental 
data~\cite{Goel2001,Liu2005}.
Eventually, \glspl{GW} can rely on cloud services to analyze the collected data 
and choose more accurate prediction models~\cite{dias2016c}.
For example, \glspl{ANN} can provide higher accuracy than other methods, but 
they may not fit sensor nodes' constraints because building an \gls{ANN} 
requires a computation intensive training phase over a large amount of data.

Additionally, the \gls{GW} can assume the responsibility of adapting sensor 
nodes' operations according to the potential savings that predictions may 
introduce.
In such cases, the \gls{GW} can estimate if it is worth to make predictions 
in sensor nodes, based on the relation between the predictions' accuracy, the 
correlation between measurements, and the error tolerated by the 
user~\cite{Jiang2011}. 
According to the expected gains, sensor nodes can be set 
to:\begin{inparaenum}[(i)] 
	\item go to sleep mode without making any measurement; 
	\item make measurements and transmit every measurement done; or
	\item make measurements, transmit them to the \gls{GW} whenever 
the prediction differs by more than an acceptance value.
\end{inparaenum}

\section{A WSN transmission model}
\label{sec:model}

\begin{figure}[t]
	\centering
	\includegraphics[width=0.5\textwidth]{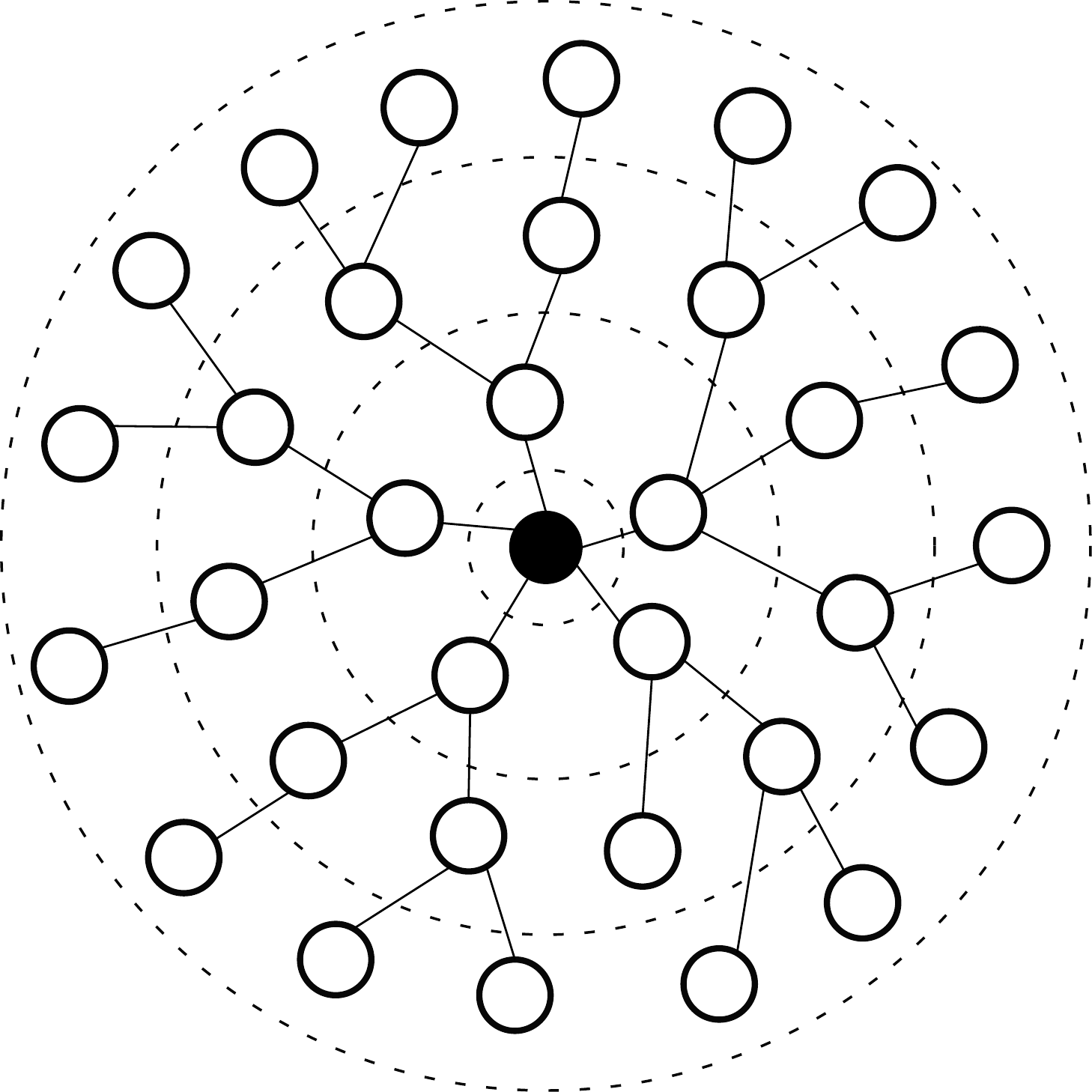}
	\caption{Sensor network model based on the density of the sensor nodes and 
their coverage. Each node has an average of five ($C = 5$) neighbors at the 
physical layer, and the vertices represent communication links established in 
an average (optimistic) scenario. The dark circle represents the 
\glsentryshort{GW}.}
	\label{fig:spanning-tree}
\end{figure}

Langendoen and Meier~\cite{Langendoen2010} presented a ring model for WSN 
topologies to describe a multi-hop network based on the average number of 
neighbors ($C$) of a sensor node and the number of hops from the 
\gls{GW} to the furthest nodes ($D$).
Assuming a uniform node density on the plane and defining it as $C + 1$ nodes 
per unit disk, the first ring ($d = 1$) is expected to have $C$ nodes. 
Figure~\ref{fig:spanning-tree} shows an example of a sensor network based on 
this model with $D = 3$ and $C = 5$.

In this model, the \gls{GW} is always in ring zero, and transmissions made 
by a component (either the \gls{GW} or a sensor node) can reach neighbors 
that are up to one unit of length from it.
The set of neighbors of a sensor node~$i$ is defined by all sensor nodes in the 
unit disk centered in $i$.
The unit disk represents the sensor node's transmission range and does not 
necessarily imply that neighbor sensor nodes will establish a direct 
communication at the routing layer.

In fact, communication links are defined by underlying routing protocols.
Langendoen and Meier assumed that these protocols aim to keep the smallest 
number of hops in a \gls{WSN} and that sensor nodes only transmit to sensor 
nodes in the previous ring, i.e., the next ring closer to the \gls{GW}.
For example, to reach the \gls{GW} from ring~$d$, we can expect a $d$-hop 
transmission.
Therefore, the distance from the \gls{GW} also defines in which ring a sensor 
node is placed. 

The expected number of sensor nodes $N_d$ in ring $d$ can be calculated based 
on the surface area of the annulus\footnote{The region bounded by two 
concentric circles.}:

\begin{equation}
N_d =  
\begin{cases} 
0, & \text{if } d = 0\\ 
Cd^2-C(d-1)^2 = (2d-1)C, & \text{otherwise}
\end{cases}
\label{eq:n_d}
\end{equation}

The number of nodes in the \gls{WSN} is equal to $CD^2$.
Given that the first ring has $C$ sensor nodes, it is expected $C$ branches 
starting in the \gls{GW} with $D^2$ sensor nodes each. 
In this work, each of these branches will be referenced as a \emph{sub-tree}.

Assuming a sensor node in ring $d$, the expected number of direct children 
($I_d$) can be calculated by the relation $N_{d+1} / N_d$.
This value does not depend on the value of~$C$:
\begin{equation}
I_d = 
\begin{cases} 
 0, & \text{if } d = D\\ 
 \frac{2d+1}{2d-1}, & \text{otherwise}
\end{cases}
\label{eq:i_d}
\end{equation}

\subsection{Model extension}

We assume that the number of transmissions and receptions made by sensor 
nodes is the primary concern in monitoring applications, not only due to the 
challenge to access the medium but also due to the energy required for the 
external communication.
Although these challenges are commonly observed in irregular real-world 
topologies, they are often neglected by other models that do not give special 
attention to the sensor nodes that may have the highest 
workloads~\cite{Deshpande2004}.
Having said that, we highlight that the main advantage of this model is its 
simplicity to identify and describe the operation of the bottlenecks in a 
sensor network, i.e., the sensor nodes in the first ring.

In this work, we will extend the original model and derive the number of 
transmissions based on the sensor nodes' positions.
First, we define the set of children nodes of a sensor node $i$ in ring $d$ as 
$H_{i,d}$.
We define the expected cardinality of $H_{i,d}$ as $K_d$. 
The value of $K_d$ is the number of direct children times the expected 
number of their children plus one (representing themselves):

\begin{equation}
 K_{d} \triangleq |H_{i,d}| =
 \begin{cases} 
  0, & \text{if } d = D\\ 
  I_{d}(K_{d+1} + 1), & \text{otherwise.}
 \end{cases}
 \label{eq:k_d}
\end{equation}
Recall that the expected number of sensor nodes is $CD^2$, and the first ring 
is expected to have $C$ nodes.
Thus, the expected number of children of the sensor nodes in the first ring 
(i.e., $K_{1}$) is $D^2 - 1$ if $D > 1$.

% \subsubsection{Volume of transmissions}

\begin{table}[t]
\centering
\def\arraystretch{1}
\rowcolors{2}{gray!25}{white}
\begin{tabular}{
>{\centering\arraybackslash}m{1.8cm}
>{\raggedright\arraybackslash}m{9.6cm}
}
{\cellcolor{gray!65}\bf Parameter} & {\cellcolor{gray!50}\bf Description} \\  
\hline 
\cellcolor{gray!15}$f$             & Number of measurements per time slot \\
\cellcolor{gray!40}$T$               & Period between the choice of two new 
prediction models \\
\cellcolor{gray!15}$C$               & Expected number of neighbors of each 
sensor node \\
\cellcolor{gray!40}$D$               & Expected number of rings/hops in the 
sensor network   \\
\cellcolor{gray!15}$\rho$             & Correlation between measurements in a 
\emph{sub-tree} \\
\cellcolor{gray!40}$\alpha$           & Expected prediction's accuracy  \\ 
\hline
\end{tabular}
\caption{Parameters taken into account to calculate the number of transmissions 
and receptions using the model.}
\label{table:model-parameters}
\end{table}

\subsubsection{Node-to-GW transmissions}

In monitoring applications, sensor nodes usually measure and transmit their 
data at a pre-defined time interval ($1/f$) that can vary from few seconds to 
hours. 
The number of measurements per second ($f$), the period between choosing 
new prediction models ($T$) and the other parameters shown in 
Table~\ref{table:model-parameters} may 
vary from case to case.

In the simplest approach, measurements are transmitted right after their 
creation.
We will assume this behavior as the baseline for further comparisons.
Alternatively, these transmissions, which we call \emph{node-to-\gls{GW}}, may 
not happen right after measurements' creation if sensor nodes aggregate the 
data received from other sensor nodes or past measurements.
The impact of aggregating transmissions will also be modeled in the following.

Given that sensor nodes must forward data from their children 
towards the \gls{GW}, the expected number of transmissions (\Ptx{_d}) during a 
period of $1/f$ seconds in a sensor node $i$ in ring $d$ is 
\begin{equation}
\Ptx{_{i,d}}~=~(K_{d} + 1)\text{,}
\label{eq:ptx_d}
\end{equation}
and the 
number of receptions is
\begin{equation}
\Prx{_{i,d}}~=~K_{d}\text{.}
\label{eq:prx_d}
\end{equation}

Finally, the total number of transmissions during a period $T$ in a sensor node 
$i$ is the sum of transmissions and receptions:
\begin{equation}
\begin{split}
E(X_{i,d}) &= \Ptx{_d} + \Prx{_d} \\
 &= ((K_{d} + 1) + K_{d})~fT \\
 &= (2K_{d} + 1)~fT\text{.}
\end{split}
\label{eq:x_d}
\end{equation}

Based on~(\ref{eq:k_d}), we can affirm that $K_{1} > K_{d}$, if $d > 1$. 
Applying this inequality to~(\ref{eq:ptx_d}), (\ref{eq:prx_d}), and 
(\ref{eq:x_d}), we mathematically show that, if $d > 1$, then $\Ptx{_{i,1}} > 
\Ptx{_{i,d}}$, $\Prx{_{i,1}} > \Prx{_{i,d}}$, and $X{_{i,1}} > X{_{i,d}}$.
Thus, in a homogeneous sensor network, sensor nodes in the first ring make more 
transmissions and are the bottlenecks that limit the number of transmissions in 
their \emph{sub-trees}, according to their capacity.
It shows the importance of focusing on optimizing the number of transmissions 
in the first ring because it can prevent new sensor nodes of being appended to 
the network.
% \footnote{In 
% Appendix~\ref{appendix:importance}, we use this model 
% to detail the importance of reducing the number of transmissions in a 
% \gls{WSN}.}.

\subsubsection{GW-to-node transmissions}

\emph{\gls{GW}-to-node} transmissions are those initiated by the 
\gls{GW}, for example, to change a configuration or update the software in the 
sensor nodes.
Assuming one unicast transmission per packet, if the \gls{GW} transmits a 
packet to every sensor node in the \gls{WSN}, a sensor node $i$ in ring $d$ 
will receive one \emph{\gls{GW}-to-node} transmission to itself, plus $K_d$  
transmissions (one per child), which must be forwarded towards their receivers.
Therefore, the number of transmissions and receptions at a sensor node $i$ in 
ring $d$ is

\begin{equation}
\Ptx{_{i,d}^*} = K_d\text{,}
\label{eq:ptx_broadcast}
\end{equation}
and
\begin{equation}
\Prx{_{i,d}^*} = K_d + 1\text{.}
\label{eq:prx_broadcast}
\end{equation}

In this case, the average number of transmissions made by the \gls{GW} to a 
\emph{sub-tree} is $D^2$, i.e., the number of nodes in each \emph{sub-tree}.
Therefore, a sensor node in the first ring will make $K_1 + 1 = D^2$ receptions 
and $K_1 = D^2 - 1$ transmissions.

\section{Modeling Dual Prediction Schemes}
\label{sec:proposed-mechanism}

As explained in Section~\ref{sec:background}, \glspl{DPS} exploit the 
proximity of the sensor nodes to the sources of the data, avoiding unnecessary 
transmissions and handling occasional sensor nodes' hardware limitations that 
might reduce \glspl{WSN}' lifetime.
A \gls{DPS} has two tasks that may be executed either by \glspl{GW} or by 
sensor nodes, namely the \emph{prediction model choice} and the 
\emph{prediction 
model dissemination}.
The \emph{dissemination} is the process of transmitting the prediction model 
either from sensor nodes to the \gls{GW} or from the \gls{GW} to sensor 
nodes.

In the following, we describe the impact of these tasks in the network load, 
concerning the number of transmissions.
Before that, we discuss the assumptions and limitations of this model.

\subsection{Assumptions and limitations}

In this model, we assume that the quality of the measurements provided by a 
\gls{WSN} can be scaled as ``acceptable'' if the values at the \gls{GW} do not 
differ by more than a certain threshold.
Since sensor nodes can compare their predictions with real measurements locally 
(without making any transmission), no transmission will be required if a
prediction is accurate, i.e., it does not differ by more than an acceptance 
threshold from the measured value.

In some cases, \glspl{WSN}' managers have no information about the 
statistics of the data that is going to be retrieved by the sensor nodes. 
Thus, it may be necessary a long \emph{initialization phase} before starting to 
make predictions. 
For example, schemes that use advanced prediction methods, like \glspl{ANN}, 
require larger amounts of data to find stable models, due to their high 
complexity and the vast number of parameters to estimate~\cite{Armstrong2001}. 
We do not include this phase in this model because we assume that the \gls{GW} 
has no energy constraints.

In this work, we do not expect distributed algorithms, i.e., sensor 
nodes do not have to synchronize with their neighbors.
However, this can be easily extended from our model, given the number of 
expected neighbors of each sensor node.

Moreover, object tracking and event detection applications are not under the 
scope of this work because they usually have other requirements than those we 
assume in this model, such as higher reliability or lower 
delays~\cite{Wang2011}.

Finally, our model is designed to represent an average \gls{WSN} with 
connectivity between its sensor nodes at the application layer.
Faulty sensors may affect the sensor node distribution, their density and force 
a new routing strategy.
Therefore, small \glspl{WSN} with low sensor nodes' density would 
require special attention, as the number of children of a sensor node may 
change and eventually overload it with extra transmissions.
On the other hand, large and dense \glspl{WSN} may not be impacted by small 
numbers of faulty sensor nodes, and our model can still be valid.

\subsection{Prediction model choice and dissemination}
\label{sec:predict-next-time-interval}

In a \gls{DPS}, the same prediction model is shared between a sensor node and 
the \gls{GW}.
Each sensor node (or group of sensor nodes) has its prediction model, and the 
prediction models in a \gls{WSN} can be independently chosen by both (sensor 
nodes and \gls{GW}) without making any new transmission.
Alternatively, prediction models can be chosen in the \gls{GW} or in sensor 
nodes.
In case that prediction models are chosen in sensor nodes, the \gls{GW} 
must receive the parameter values and, in some cases, also the prediction 
method selected.
On the other hand, if the \gls{GW} is responsible for choosing prediction 
models, sensor nodes must be informed about the decisions taken.

Assuming that the \emph{dissemination} of a prediction model is made through a 
unicast transmission, sensor nodes in the first ring will receive and 
forward every transmission to their children towards the proper destinations.
Thus, if the \gls{GW} is responsible for generating the prediction models, 
the sensor nodes in the first ring will have to forward the transmissions 
to their children.
In such cases, a sensor node in the first ring will receive $D^2$ packets.
From these packets, $D^2 - 1$ will be forwarded to its children.
Therefore, to disseminate the prediction models generated in the \gls{GW}, 
there will be 

\begin{equation}
\begin{split}
\Xtop{-GW} &= \Prx{_{i,1}^*} + \Ptx{_{i,1}^*} \\
	&= D^2 + (D^2 - 1) \\
	&= 2D^2 - 1
	\label{eq:n-dis-gw}
\end{split}
\end{equation}
transmissions (summing transmissions and receptions) in each sensor node in the 
first ring.

Analogously, in case that prediction models are chosen in the sensor nodes, 
every sensor node in the first ring will make $D^2$ transmissions to the 
\gls{GW} after receiving $D^2 - 1$ prediction models.
Thus, the number of transmissions at the first ring will be, once again, 
equal to $2D^2 - 1$.

If packets to the same \emph{sub-tree} are aggregated, sensor nodes in 
the first ring will 
receive only one packet that will be split before being retransmitted to
their direct children in the second ring. 
In such cases, a sensor node in the first ring will need  

\begin{equation}
\Xtop{-GW-AGG} = \Xtop{-SN-AGG} = I_{1} + 1
\label{eq:n-dis-gw-aggregation}
\end{equation}
transmissions to \emph{disseminate} the prediction models, where, 
from~(\ref{eq:i_d}), $I_{1} = 3$, if $D > 1$.

Finally, if the \gls{GW} uses broadcast (or multicast) transmissions,
sensor nodes will receive and forward only one packet, i.e.,
\begin{equation}
\Xtop{-GW-BC} = 1\text{.}
\label{eq:n-dis-gw-broadcast}
\end{equation}

\subsection{Impact of predictions in the number of transmissions}

As described before, adopting a data prediction scheme can benefit the \gls{WSN}
reducing the number of transmissions and optimizing the medium access control, 
which may eventually reduce energy consumption and extend the \gls{WSN}'s 
lifetime.
To estimate the number of transmissions in homogeneous networks, we develop a 
formula based on the predictions' accuracy and the correlation of the monitored 
data.

Let us assume that $\alpha_{i}$ is the accuracy of the predictions in sensor 
node $i$, i.e., $\alpha_{i}$ is the probability that a measurement of $i$ 
matches to the prediction and does not have to be transmitted to the \gls{GW}, 
and $\alpha_{i}^c = 1 - \alpha_{i}$. 
Therefore, the expected number of transmissions and receptions
in a sensor node $i$ during a time interval of $1 / f$ seconds (i.e., between 
two measurements) is respectively represented by $\Ptx{'_{i,d}}$ and 
$\Prx{'_{i,d}}$ as
\begin{equation}
\Ptx{'_{i,d}} = \alpha^c_i + \sum_{j \in H_{i,d}}~\alpha^c_j \text{,}
\label{eq:s-line}
\end{equation}
and

\begin{equation}
\Prx{'_{i,d}} = \sum_{j \in H_{i,d}}~\alpha^c_j \text{.}
\label{eq:r-line}
\end{equation}

Considering an eventual \emph{dissemination} of the prediction models,
the expected number of transmissions and receptions
during a period of $T$ seconds is

\begin{equation}
\begin{split}
 E(X'_{i,d}) &= (\Ptx{'_{i,d}} + \Prx{'_{i,d}})\frequency{}T + \Xtop{} \\
&= (\alpha^c_i + \sum_{j \in H_{i,d}}~\alpha^c_j + \sum_{j \in 
H_{i,d}}~\alpha^c_i)\frequency{}T + \Xtop{} \\
 &= (\alpha^c_i + 2\sum_{j \in H_{i,d}}~\alpha^c_j)\frequency{}T + \Xtop{}
 \end{split}
 \label{eq:x-line}
\end{equation}

Notice that a low accuracy in predictions used in sensor nodes that are far 
from 
the \gls{GW} has a higher impact on the total number of \gls{WSN} transmissions 
than a low accuracy in predictions used in sensor nodes in the first rings.
However, concerning the number of transmissions at a single sensor node, the 
bottleneck of the \gls{WSN} is still represented by sensor nodes in the first 
ring.

Let us define the minimum average accuracy ($\alpha_\text{min}$) necessary
to reduce the number of transmissions, according to the size of the network and 
its number of rings.
This value can be used to define the maximum number of transmissions 
($\Ptxmax{'_{i,d}}$) and receptions ($\Prxmax{'_{i,d}}$)
in a sensor node $i$ in ring$~d$:
\begin{equation}
\begin{split}
\Ptxmax{'_{i,d}} 	&= (1 - \alpha_\text{min}) + \sum_{j \in 
H_{i,d}}~(1 - \alpha_\text{min}) \\
		&= (1 + K_{d})~(1 - \alpha_\text{min})
\end{split}
\label{eq:s-line-max}
\end{equation}
and

\begin{equation}
\begin{split}
\Prxmax{'_{i,d}} 	&= \sum_{j \in H_{i,d}}~(1 - \alpha_\text{min}) \\
		&= K_{d}~(1 - \alpha_\text{min})
\end{split}
\label{eq:r-line-max}
\end{equation}

Recall that $K_{d} \triangleq |H_{i,d}|$, for a sensor node $i$ in ring $d$.
Therefore, 
\begin{equation}
E(X'_{i,d}) \leq ((1 + K_{d})~(1 - \alpha_\text{min}) + K_{d}~(1 - 
\alpha_\text{min}))  fT + \Xtop{}
\label{eq:n-line}
\end{equation}

Finally, the use of predictions will reduce the number of transmissions 
if $X_{i,d}' < X_{i,d}$.
After some mathematical development shown in~\ref{appendix:accuracy}, 
we arrive at the following expression for the minimum average accuracy of the 
predictions:

\begin{equation}
\begin{split}
\alpha_\text{min} &> \frac{\Xtop{}}{(2D^2  - 1) f T}
\end{split}
\end{equation}

In conclusion, if prediction models are not independently generated
in \glspl{GW} and sensor nodes, the DPS requires a minimum accuracy to 
ensure the reduction in the number of transmissions. 
Hence, the minimum accuracy is a lower bound that depends only on the network 
layout (i.e., the number of rings $D$), the frequency of the measurements 
(\frequency{}) and the time between choosing two prediction models ($T$).
If the predictions' accuracy does not reach this limit, there will exist 
two actions to improve the network operation, either to set new values for 
$f$ and $T$ or to turn the \gls{DPS} off.

\subsection{Impact of predictions in the number of transmissions}

As described before, adopting a data prediction scheme can benefit the \gls{WSN}
reducing the number of transmissions and optimizing the medium access control, 
which may eventually reduce energy consumption and extend the \gls{WSN}'s 
lifetime.
To estimate the number of transmissions in homogeneous networks, we develop a 
formula based on the predictions' accuracy and the correlation of the monitored 
data.

Let us assume that $\alpha_{i}$ is the accuracy of the predictions in sensor 
node $i$, i.e., $\alpha_{i}$ is the probability that a measurement of $i$ 
matches to the prediction and does not have to be transmitted to the \gls{GW}, 
and $\alpha_{i}^c = 1 - \alpha_{i}$. 
Therefore, the expected number of transmissions and receptions
in a sensor node $i$ during a time interval of $1 / f$ seconds (i.e., between 
two measurements) is respectively represented by $\Ptx{'_{i,d}}$ and 
$\Prx{'_{i,d}}$ as

\begin{equation}
\Ptx{'_{i,d}} = \alpha^c_i + \sum_{j \in H_{i,d}}~\alpha^c_j \text{,}
\label{eq:s-line}
\end{equation}
and

\begin{equation}
\Prx{'_{i,d}} = \sum_{j \in H_{i,d}}~\alpha^c_j \text{.}
\label{eq:r-line}
\end{equation}

Considering an eventual \emph{dissemination} of the prediction models,
the expected number of transmissions and receptions
during a period of $T$ seconds is

\begin{equation}
\begin{split}
 E(X'_{i,d}) &= (\Ptx{'_{i,d}} + \Prx{'_{i,d}})\frequency{}T + \Xtop{} \\
&= (\alpha^c_i + \sum_{j \in H_{i,d}}~\alpha^c_j + \sum_{j \in 
H_{i,d}}~\alpha^c_i)\frequency{}T + \Xtop{} \\
 &= (\alpha^c_i + 2\sum_{j \in H_{i,d}}~\alpha^c_j)\frequency{}T + \Xtop{}
 \end{split}
 \label{eq:x-line}
\end{equation}

Note that a low accuracy in predictions used in sensor nodes that are far from 
the \gls{GW} has a higher impact on the total number of \gls{WSN} transmissions 
than a low accuracy in predictions used in sensor nodes in the first rings.
However, concerning the number of transmissions at a single sensor node, the 
bottleneck of the \gls{WSN} is still represented by the sensor nodes in the 
first 
ring.

Let us define the minimum average accuracy ($\alpha_\text{min}$) necessary
to reduce the number of transmissions, according to the size of the network and 
its number of rings.
This value can be used to define the maximum number of transmissions 
($\Ptxmax{'_{i,d}}$) and receptions ($\Prxmax{'_{i,d}}$)
in a sensor node $i$ in ring$~d$:
\begin{equation}
\begin{split}
\Ptxmax{'_{i,d}} 	&= (1 - \alpha_\text{min}) + \sum_{j \in 
H_{i,d}}~(1 - \alpha_\text{min}) \\
		&= (1 + K_{d})~(1 - \alpha_\text{min})
\end{split}
\label{eq:s-line-max}
\end{equation}
and

\begin{equation}
\begin{split}
\Prxmax{'_{i,d}} 	&= \sum_{j \in H_{i,d}}~(1 - \alpha_\text{min}) \\
		&= K_{d}~(1 - \alpha_\text{min})
\end{split}
\label{eq:r-line-max}
\end{equation}

Recall that $K_{d} \triangleq |H_{i,d}|$, for a sensor node $i$ in ring $d$.
Therefore, 
\begin{equation}
E(X'_{i,d}) \leq ((1 + K_{d})~(1 - \alpha_\text{min}) + K_{d}~(1 - 
\alpha_\text{min}))  fT + \Xtop{}
\label{eq:n-line}
\end{equation}

Finally, the use of predictions will reduce the number of transmissions 
if $X_{i,d}' < X_{i,d}$.
After some mathematical development shown in~\ref{appendix:accuracy}, 
we arrive at the following expression for the minimum average accuracy of the 
predictions:

\begin{equation}
\begin{split}
\alpha_\text{min} &> \frac{\Xtop{}}{(2D^2  - 1) f T}
\end{split}
\end{equation}

In conclusion, intuitively, if prediction models are independently generated in 
\glspl{GW} and sensor nodes, the \gls{DPS} does not require a minimum accuracy 
to ensure the reduction in the number of transmissions. 
Otherwise, if prediction models are not independently generated in \glspl{GW} 
and sensor nodes, the extra transmissions used to disseminate new prediction 
models may turn the prediction scheme into an inefficient option.
Hence, the efficiency of a \gls{DPS} also depends on how many transmissions are 
required for disseminating the prediction models because the number of 
transmissions will be proportional to the number of hops between sensor nodes 
and the \gls{GW}.

Moreover, the minimum accuracy is a lower bound that depends only on the 
network layout (i.e., the number of rings $D$), the frequency of the 
measurements (\frequency{}) and the time between two predictions ($T$).
Therefore, if the predictions' accuracy does not reach this limit, there will 
exist three actions to improve the network operation, either to set new values 
for $f$ and $T$, to adopt a \gls{DPS} with independent model generation, 
or to turn the \gls{DPS} off.

\subsection{Impact of predictions and data aggregation}

Additionally to \glspl{DPS}, it may be possible to adopt aggregation schemes in 
sensor nodes, such that a sensor node aggregates data received from its 
children 
and transmits only after making its measurement.
In the following, we model an aggregation scheme and compare its efficiency 
with the use of \glspl{DPS}.
Finally, we estimate the reduction in the number of transmissions if both 
techniques are simultaneously adopted.

To make it clear for the reader, we introduce a scenario with only two sensor 
nodes to clarify the normalization of the data and its application.
Later, we will extend the model to a more complex scenario with $D$ rings.

\subsubsection{Network with two sensor nodes}

\begin{figure}[t]
	\centering
	\includegraphics[width=0.5\textwidth]{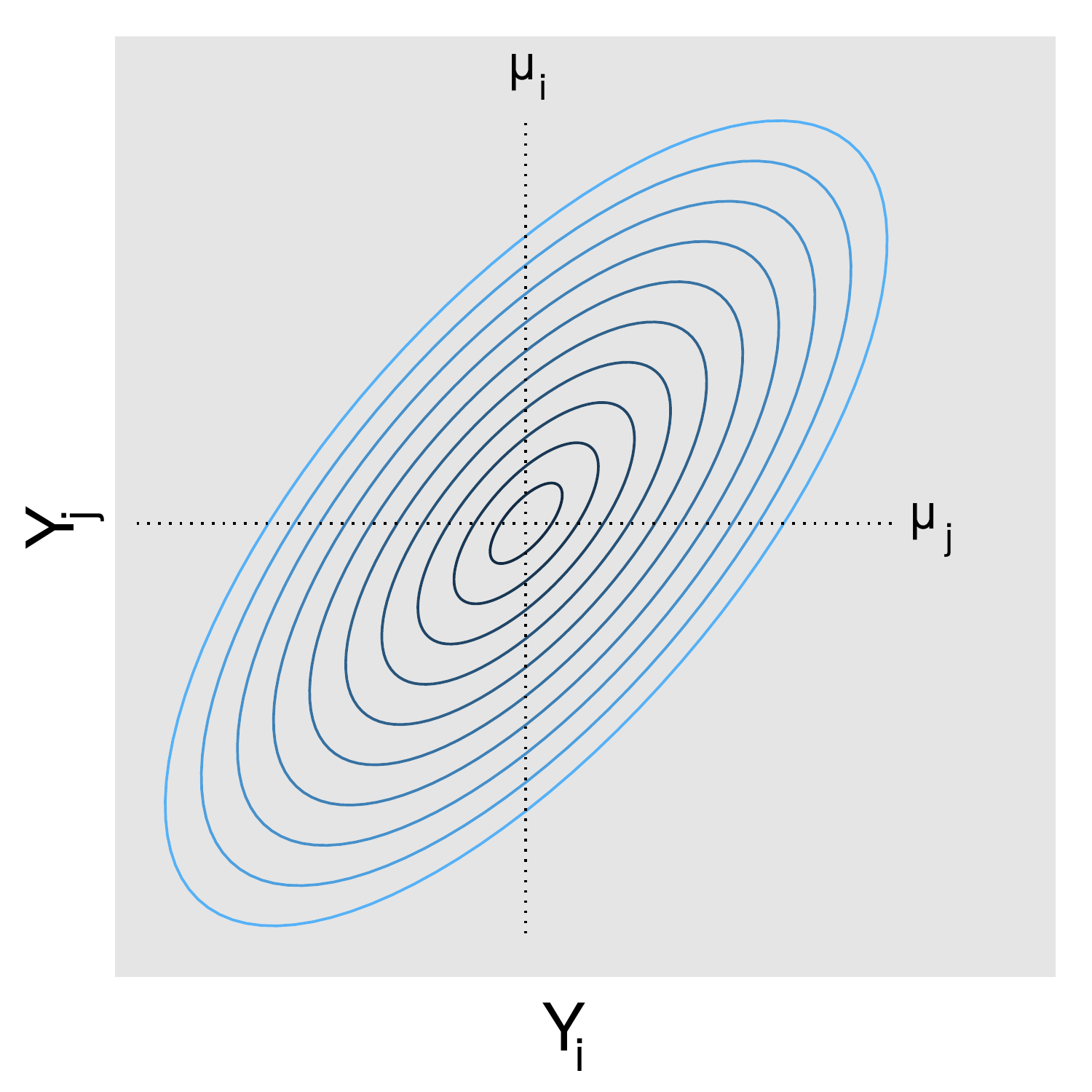}
	\caption{Values of $Y_i$ and $Y_j$ are correlated ($\rho_{i,j} = 0.7$), 
and each line represents a different density of points.}
	\label{fig:ellipse}
\end{figure}

Let us consider a section of the sensor network with the \gls{GW} and a 
sensor node $i$ with a single child $j$. 
Due to the sensor network's topology, transmissions from sensor node $j$ can 
reach the \gls{GW} only through $i$.
Thus, every $1/\frequency{}$ seconds,$~i$ may transmit to the \gls{GW} if its 
prediction has failed or if it had happened to $j$.

We assume that the measurements of $i$ and $j$ follow the Normal distributions 
respectively represented by $Y_i = N(\mu_i, \sigma^2_i)$ and $Y_j = N(\mu_j, 
\sigma^2_j)$. 
Such a \gls{MVN} distribution can be defined based on the 
correlation between their values, i.e., the relationship between each 
pair of measurements made by $i$ and $j$.
An illustration of the \gls{MVN} density containing $Y_i$ and $Y_j$ is 
shown in Figure~\ref{fig:ellipse}.

Assuming that the predictions ($\bar{\text{y}}$) are not biased 
(i.e., $\bar{\text{y}} = \mu$), 
we may also approximate them to 
Normal distributions\footnote{In~\ref{appendix:data}, we detail how 
to estimate the predictions' accuracy for normally distributed measurements, 
based on the user's acceptance threshold for errors.} and label an outcome as 
incorrect whenever a measurement falls outside the interval defined by the 
acceptance threshold $\varepsilon$.
In such cases, the probability that the sensor node~$j$ will transmit 
(including its measurement) after $1/\frequency{}$ seconds is 
$1~-~\alpha_j$.
Hence, the probability of $i$ receiving a packet is also $1~-~\alpha_j$.

Similarly, $i$ will transmit if the prediction about its measurement fails 
(i.e., it differs from the actual measurement more than the acceptance 
threshold $\varepsilon_i$) or if the prediction in sensor node $j$ 
had failed and a measurement has been received.
In other words, there will be a transmission if at least one of the 
two predictions fail.

\begin{figure}[t]
	\centering
	\includegraphics[width=0.5\textwidth]{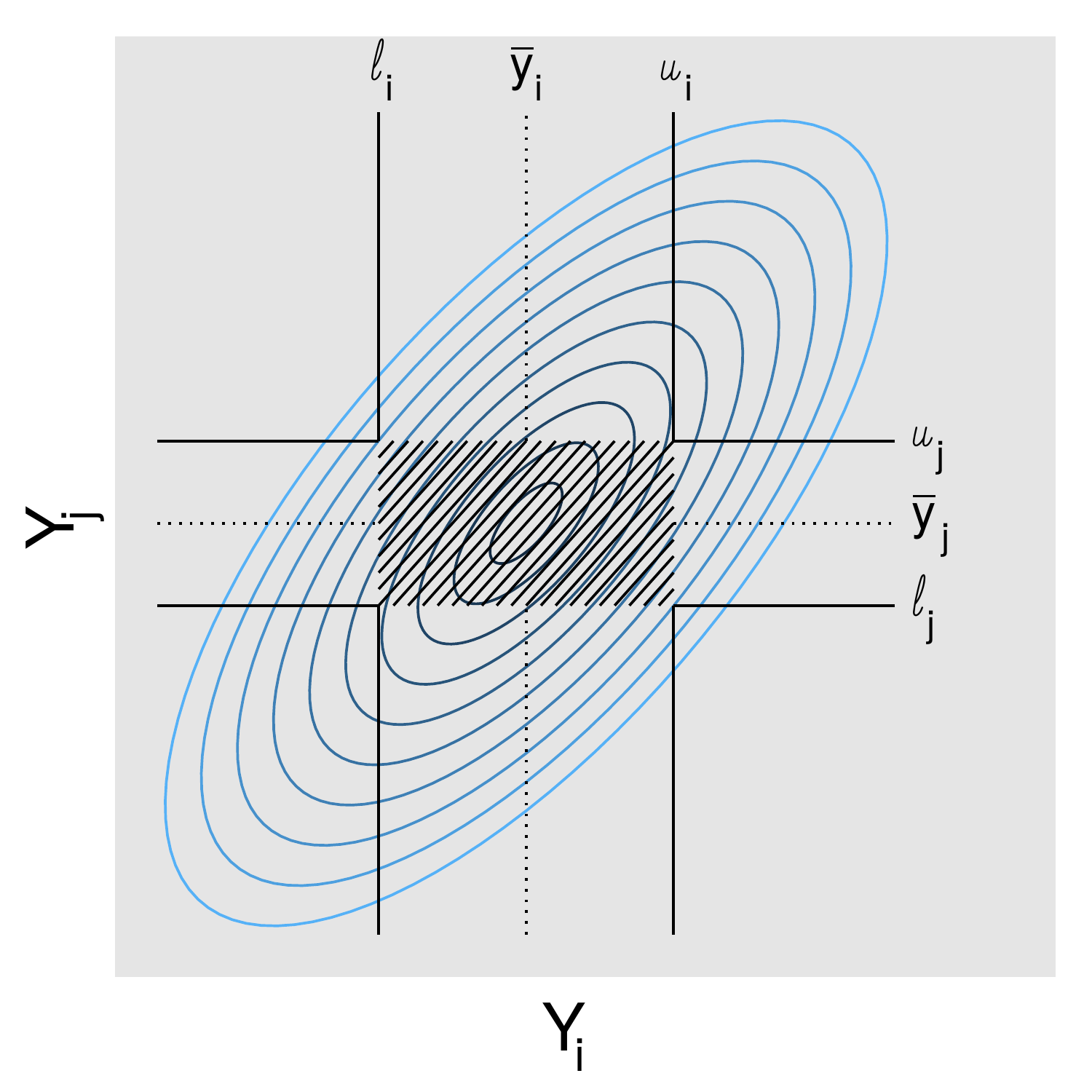}
	\caption{The hashed rectangle in the center illustrates the points in 
which both predictions ($\bar{\text{y}}_i$ and $\bar{\text{y}}_j$) are correct.}
	\label{fig:ellipse-square}
\end{figure}

If $i$ can aggregate transmissions, its total number of transmissions is not a 
simple sum as in the case without aggregation because it depends on the 
correlation of the measurements of $i$ and $j$.
Let us assume that the correlation between $Y_i$ and $Y_j$ is defined by the 
Pearson correlation coefficient and represented by $\rho_{i,j}$. Therefore, to 
model the probability of having at least one wrong prediction, we must 
calculate the correlation matrix ($\Sigma$), which is defined as

\begin{equation}
	\Sigma = \begin{bmatrix} 
			\sigma_{i}^2 & \rho_{i,j}~\sigma_{i}\sigma_{j} \\  
			\rho_{i,j}~\sigma_{i}\sigma_{j} &  \sigma_{j}^2 
		\end{bmatrix}
\end{equation}
Finally, given the lower limits

\begin{equation}
l_i = \bar{y}_i - \varepsilon_i \text{ and } l_j = \bar{y}_j - \varepsilon_j,
\end{equation}
the upper limits
\begin{equation}
u_i = \bar{y}_i + \varepsilon_i \text{ and } u_j = \bar{y}_j + \varepsilon_j,
\end{equation}
and the correlation matrix ($\Sigma$), it is possible to 
calculate the following \gls{MVN} probability:

\begin{equation}
F(y_i, y_j) = \frac{1}{\sqrt{|\Sigma| (2\pi)^2}} 
	\int_{l_i}^{u_i} 
	\int_{l_j}^{u_j}
	e^{\left(- \frac{1}{2}\theta^t \Sigma^{-1} \theta\right)} d\theta
\end{equation}
The value of $F(y_i, y_j)$ represents the probability that both 
predictions (in $i$ and $j$) are correct and can be illustrated by the density 
inside the crosshatched rectangle in Figure~\ref{fig:ellipse-square}.
Thus, the probability of at least one prediction failing can be 
calculated as $(1~-~F(y_i, y_j))$, which, in fact, is the probability of sensor 
node $i$ making a transmission after $1/\frequency{}$ seconds. 

Finally, considering occasional extra transmissions to disseminate the 
prediction 
model, 
the number of transmissions at $j$ during a period of $T$ seconds can be 
calculated as $(1~-~\alpha_j) \frequency{}T + \Xtop{}$, and
the expected sum of transmissions and receptions at sensor node 
$i$ during a period of $T$ seconds can be modeled as $((1~-~F(y_i, 
y_j))+(1~-~\alpha_j)) \frequency{}T+ \Xtop{}$.

\subsubsection{Larger networks}

Now, we will extend the previous example to larger sensor networks.
The correlation matrix ($\Sigma$) of several data distributions can be 
calculated as 

\begin{equation}
	\Sigma = \begin{bmatrix} 
			\sigma_{a}^2 & \rho_{a,b}~\sigma_{a}\sigma_{b} & 
\cdots & \rho_{a,z}~\sigma_{a}\sigma_{z} \\  
			\rho_{b, a}~\sigma_{b}\sigma_{a} & 
\sigma^2_{b} & \cdots & \rho_{b,z}~\sigma_{b}\sigma_{z} \\  
			\vdots & \vdots & \ddots & \vdots \\  
\rho_{z, a}~\sigma_{z}\sigma_{a} & \rho_{z,b}~\sigma_{z}\sigma_{b} & \cdots &  
\sigma_{z}^2 
		\end{bmatrix},
\end{equation}
and, similarly to the two-dimensional model, the expected number of 
transmissions made by sensor node $i$ in ring $d$ (represented by 
\Ptx{''_{i,d}}) depends not only on its predictions but also on the predictions 
used in all of its children.
The value of \Ptx{''_{i,d}} can be calculated as

\begin{equation}
\Ptx{''_{i,d}} = 1 - F(i,a,b,\ldots,z),
\end{equation}
where $\{a, b, \ldots, z \in H_{i,d}\}$, and the function $F$ is the 
\gls{MVN} probability function integrated from the lower accepted 
limits to the upper accepted limits over the $k = 1 + K_{d}$ distributions:

\begin{equation}
F(i,a,b,\ldots,z) = \frac{1}{\sqrt{|\Sigma| (2\pi)^k}} 
	\int_{l_i}^{u_i} 
	\int_{l_a}^{u_a} 
	\int_{l_b}^{u_b}
	\cdots
	\int_{l_z}^{u_z}
	e^{\left(- \frac{1}{2}\theta^t \Sigma^{-1} \theta\right)} d\theta\text{,}
\label{eq:f-a-b-z}
\end{equation}
which can be efficiently calculated with 
the use of Monte Carlo methods for higher dimensions~\cite{Genz1993}.

The number of receptions at $i$ (\Prx{''_{i,d}}) is slightly different from the 
previous example, since now the sensor node may have several children in the 
next ring, and their transmissions happen independently.
Let us define $H'_{i,d}$ as the set of direct children of $i$.
Thus, $|H'_{i,d}| \triangleq I_d$. 
The expected number of receptions can be calculated as

\begin{equation}
\Prx{''_{i,d}} = \sum_{j \in H'_{i,d}} \Ptx{''_{j,d+1}},
\end{equation}
and the total number of transmissions and receptions is expected to be 

\begin{equation}
E(X''_{i,d})~=~(\Ptx{''_{i,d}}+\Prx{''_{i,d}}) \frequency{}~T + \Xtop{}\text{.}
\end{equation}

Even though the function $F$ has no closed formula, it is possible to set a 
lower bound based on a case when there is absolutely no correlation between the 
values measured by $i$ and its children. 
When the correlation is equal to zero, the expected number of transmissions 
and receptions at sensor node $i$ are the maximum possible.
Considering that there will exist a transmission if at least one prediction 
fails, the probability of having no transmissions at $i$ is $\alpha^{1 + 
K_{d}}$.
Thus,
\begin{equation}
\Ptxmax{''_{i,d}} = 1 - \alpha^{1 + K_{d}}\text{.}
\end{equation}

Recall that $i$ is expected to have $I_d$ direct children and each child 
be part of a \emph{sub-tree} with $K_{d} / I_d$ sensor nodes.
There may exist $I_d$ independent receptions at $i$, and each reception may 
not occur with probability $\alpha^{K_{d} / I_d}$.
Thus,
\begin{equation}
\Prxmax{''_{i,d}} = I_d~(1 - \alpha^{K_{d} / I_d})\text{.}
\end{equation}

Therefore, 
\begin{sloppypar}
\begin{equation}
\begin{split}
E(X''_{i,d}) \leq ~[~(1 - \alpha^{1 + K_{d}}) + I_d~(1 - 
\alpha^{K_{d} / I_d})~] \frequency{}~T ~~+ \Xtop{}\text{.}
\label{eq:n-line-line}
\end{split}
\end{equation}

We claim that $E(X''_{i,d}) \leq E(X'_{i,d})$, which means that a mechanism that 
aggregates the data will not make more transmissions than the one that only 
makes predictions. 
Comparing~(\ref{eq:n-line-line}) with~(\ref{eq:n-line}), we have that for any 
$\alpha \in [0, 1]$ and $K_{d} \geq 0$, it can be shown\footnote{Based on the 
proof detailed in~\ref{appendix:inequality}.} that
${(1 - \alpha^{1 + K_{d}}) \leq ((1 + K_{d}) (1 - \alpha))}$ and, hence,
$\Ptxmax{''_{i,d}} \leq \Ptxmax{'_{i,d}}$.
Moreover, 
$\Prxmax{''_{i,d}} \leq \Prxmax{'_{i,d}}$
and $I_d~(1 - \alpha^{K_{d} / I_d}) \leq K_{d}~(1 - \alpha)$,
which can be similarly proved to be true, since $(K_{d} / I_{d}) \geq 1$ when 
$K_d > 0$ and $\alpha \in [0, 1]$. In the case of being in the last ring, 
since there are no children ($K_{d} = I_{d} = 0$), no reception is made.
\end{sloppypar}

\section{Model experimentation}
\label{sec:experimentation}

Using the model presented before, we can estimate the effects of adopting a 
prediction or an aggregation scheme in a sensor network, concerning the number 
of transmissions and, eventually, the energy consumption levels.
In this Section, we make a parameter study over the model parameters $C$, $D$, 
$\rho$, and $\alpha$.
Our goal is to observe how the number of transmissions varies in different 
scenarios and apply the model using simulations with normally distributed 
data.

\subsection{Simulation setup}

In OMNET++~\cite{Varga2001}, we simulated TelosB motes~\cite{Platform} using a 
\glsentryshort{TDMA}-based \gls{MAC} protocol.
In the \gls{MAC} protocol adopted, each sensor node has a reserved slot to 
transmit. 
Therefore, we did not experience collisions during the transmissions, and there 
was no overhearing. 
We highlight that other \gls{MAC} protocols may obtain different results, due 
to concurrent transmissions, although we can expect a similar reduction
in their number of transmissions.

Regarding the mechanisms adopted to reduce the number of transmissions, we 
simulated three combinations:\begin{inparaenum}[(i)]
 \item with no prediction and no aggregation;
 \item with prediction, but no aggregation; and
 \item with aggregation, but no prediction.
\end{inparaenum}
When data prediction was adopted, the prediction models were chosen in the 
\gls{GW}, and \emph{\gls{GW}-to-node} transmissions were always aggregated.

As we showed before, in monitoring applications with \glspl{DPS}, the number 
of transmissions is highly affected by the correlation between measurements 
made 
by the sensor nodes in a \emph{sub-tree}, and by the predictions' accuracy.
Therefore, regarding the model parameters, we observed the impact of different 
values of $\rho$, $\alpha$.
Values of $\rho$ varied among $0.1, 0.2, \ldots, 0.9, \text{ and } 
0.95$, and values $\alpha$ varied among $0.5, 0.7, 0.9, \text{ and } 0.95$.

Recall that, according to~(\ref{eq:n-line-line}), the number of transmissions 
does not depend on the density of sensor nodes ($C$), but on the number of 
rings ($D$).
Thus, to observe the impact of the growth in the number of sensor nodes in 
\glspl{WSN}, we observed the number of transmissions with values of $D$ 
varying among 
$1, 2, \ldots,~\text{and}~10$.
Note that, when new rings are added, the number of sensor nodes increases 
quadratically if no aggregation is adopted.
However, the number of transmissions does not change if sensor nodes aggregate 
them.
% \footnote{In Appendix~\ref{appendix:importance}, we show that, if the 
% aggregation is not adopted, the number of transmissions increases cubically.}.

Finally, in our simulations, sensor nodes made one measurement per minute ($f = 
1/60$), and the \gls{GW} predicted their measurements once a day during three 
days ($T = 3 \times 86400$ seconds). 
Therefore, each sensor node made $4320$ measurements, from which $1440$ 
happened between each prediction model choice (in the cases
when predictions were adopted).

\subsection{Simulated algorithm}
\label{sec:simulated-algorithm}

\begin{algorithm}[t]
	\KwData{$n =$ number of nodes, $\alpha =$ accuracy, $\rho =$ 
correlation}
	\KwResult{$P (n, \alpha, \rho) =$ probability that no transmission 
happens }

	\eIf{$n = 0$}{
		\Return $P \gets 1$
	}{
		$q \gets \mid \Phi^{-1}\left(\frac{1 - \alpha}{2}\right)|$ \\
		$Q \gets \{q, q, \ldots, q\}_{1 \times n}$ \\
		$Y \gets \{Y_1, Y_2, \ldots, Y_n\}$\\
		$\Sigma \gets \begin{bmatrix} 
			1 & \rho & \cdots & \rho \\  
			\rho & 1 & \cdots & \rho \\  
			\vdots & \vdots & \ddots & \vdots \\  
			\rho & \rho & \cdots &  1 
		\end{bmatrix}_{n \times n}$
		\newline \\
		\Return $P \gets \Phi \left( Y, \Sigma, Q\right)$
	}
	\caption{Algorithm to calculate the probability that no transmission 
will be made.}
	\label{algorithm:probability-no-transmission}
\end{algorithm}

Assuming normally distributed values, the expected number of transmissions and 
receptions can be estimated using the cumulative density functions of \gls{MVN} 
distributions. 
Based on~(\ref{eq:f-a-b-z}), we designed the algorithm described in 
Algorithm~\ref{algorithm:probability-no-transmission}.
It calculates the probability $P$ of making no transmissions in a group of 
$n$ sensor nodes measuring data with correlation $\rho$ if the average 
predictions' accuracy is $\alpha$.

We highlight that, in our model, the number of children is used to define how 
many distributions will be used, which means that decimal values cannot be 
considered.
Hence, we rounded all of them up to the next integer, which resulted on 
an upper bound for the number of transmissions in the simulations.

\afterpage{
\begin{figure}[!t]
        \centering
	\begin{subfigure}[t]{0.98\textwidth}
		\includegraphics[width=\textwidth]{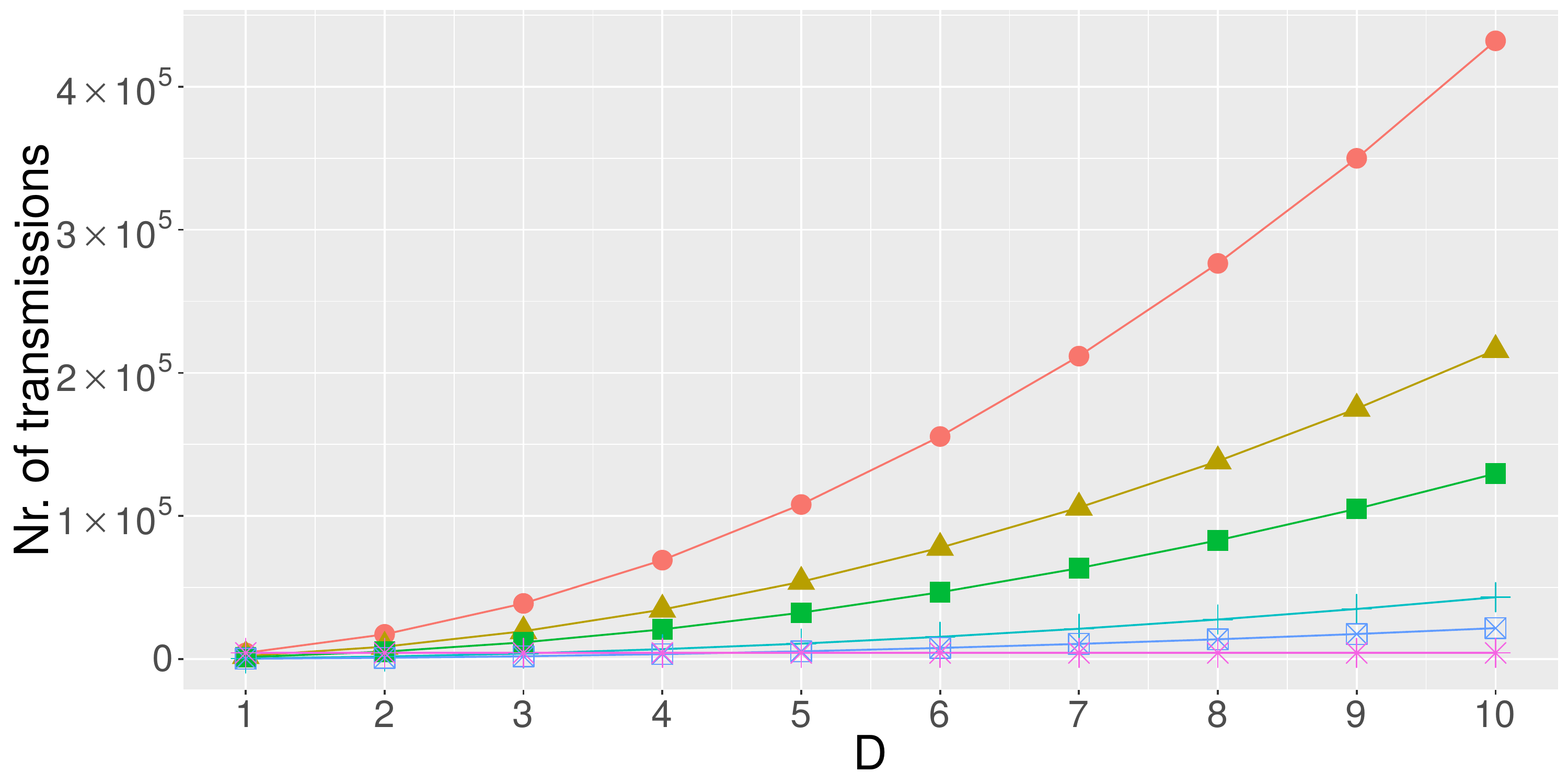}
		\caption{The aggregation reduces the number of transmissions 
from quadratic to linear order.}
		\label{fig:tx-predictions-only}
	\end{subfigure}
	\qquad
\begin{subfigure}[t]{0.49\textwidth}
\includegraphics[width=\textwidth]{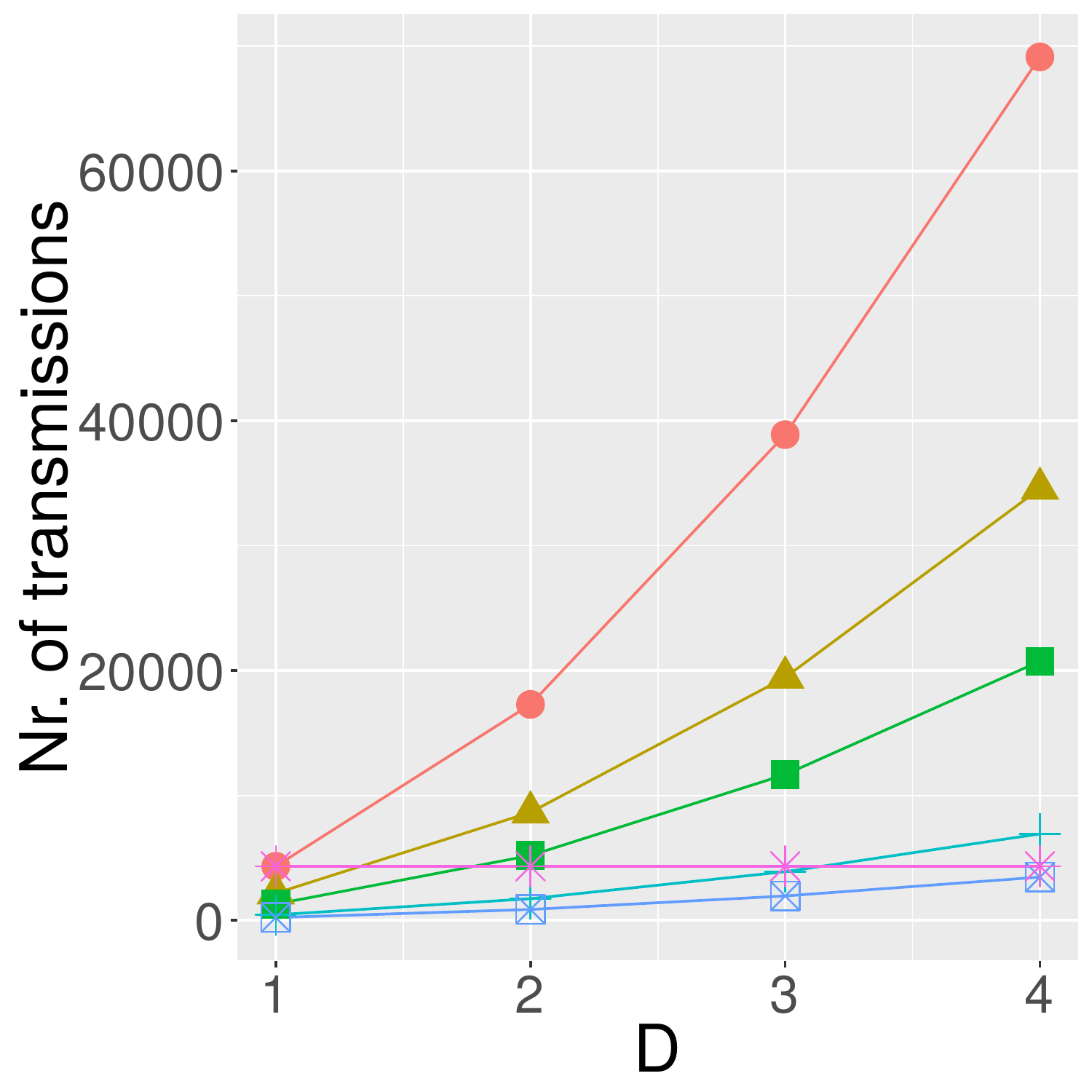}
		\caption{When number of rings is small ($D~\leq~4$), the 
use of predictions can lead to fewer transmissions than the aggregation scheme.}
		\label{fig:tx-predictions-only-highlight}
	\end{subfigure}
        \qquad
	\begin{subfigure}[t]{0.43\textwidth}
		\includegraphics[width=\textwidth]{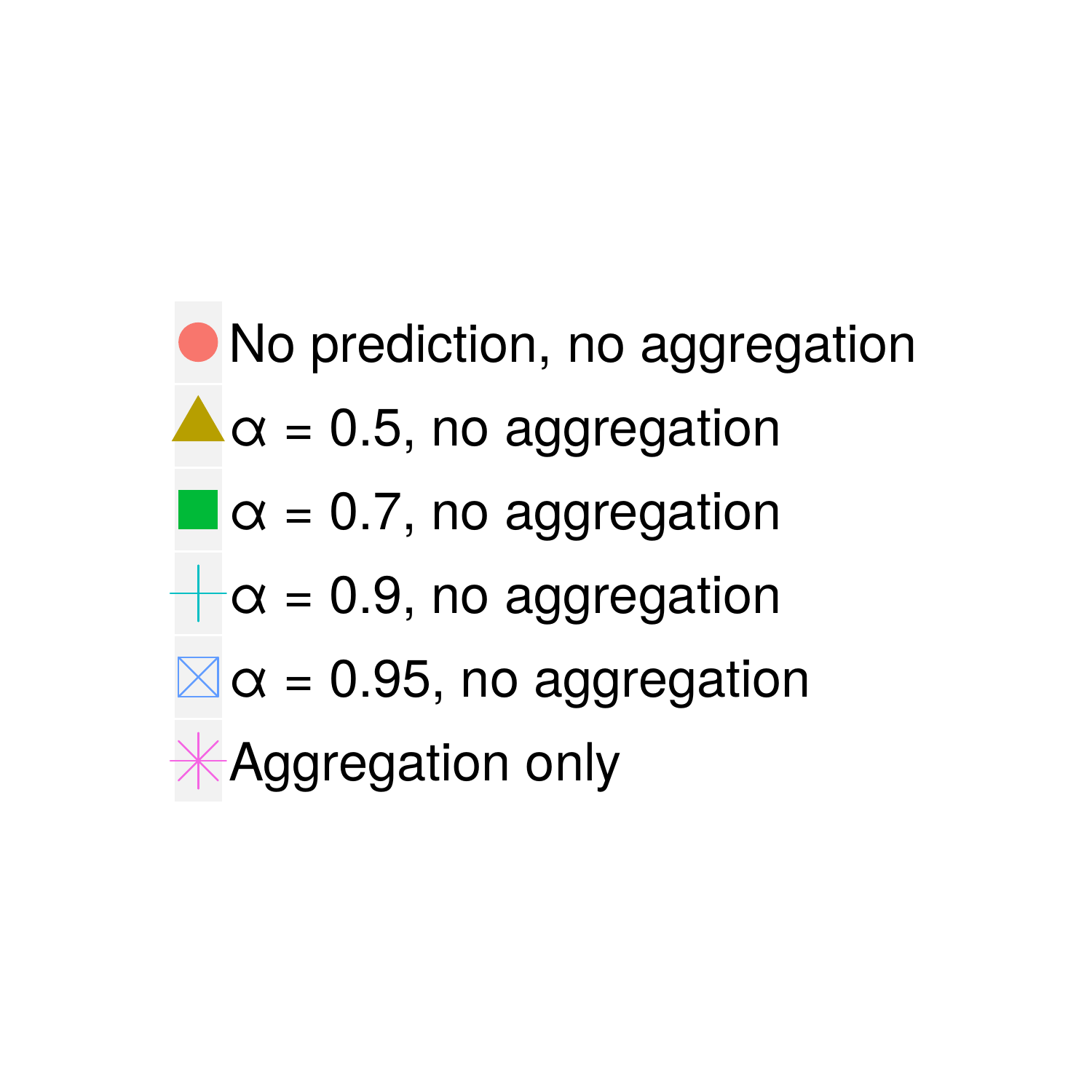}
		\label{fig:tx-predictions-only-legend}
	\end{subfigure}
	\caption{The impact of the network size in the number of transmissions in 
the first ring.}
        \label{fig:wsn}
\end{figure}
\clearpage
}

\subsection{Number of transmissions}

Given that the bottlenecks of a sensor network are the sensor nodes in the 
first 
ring, we calculate the number of transmissions at a sensor node $i$ in ring 
$d=1$ 
as

\begin{equation}
\Ptx{''_{i,1}} = (~(~(1 - P(K_1, \alpha, \rho)~)~f~) + I_1~)~T\text{,}
\end{equation}
and the number of receptions as

\begin{equation}
\Prx{''_{i,1}} = (1 - P(K_1, \alpha, \rho)~)~I_1~f~T\text{.}
\end{equation}

Figure~\ref{fig:wsn} shows the results for all the tested configurations.
In larger sensor networks ($D > 4$), data aggregation has a higher 
impact than data prediction in the number of transmissions, as shown in 
Figure~\ref{fig:tx-predictions-only}.
Similar results were observed in another study~\cite{Santini2006}, 
but the authors did not realize that the predictions had less impact in the 
final savings and concluded that such optimal achievements happened due to the 
high accuracy of the predictions.

When the predictions are highly accurate, and the number of rings is small ($D 
\leq 4$), the data prediction has a higher impact on the number of 
transmissions if compared with the scenarios where the data is only aggregated.
Figure~\ref{fig:tx-predictions-only-highlight} highlights scenarios with less 
than five rings.

\begin{figure}[t]
	\centering
	\includegraphics[width=0.98\textwidth]{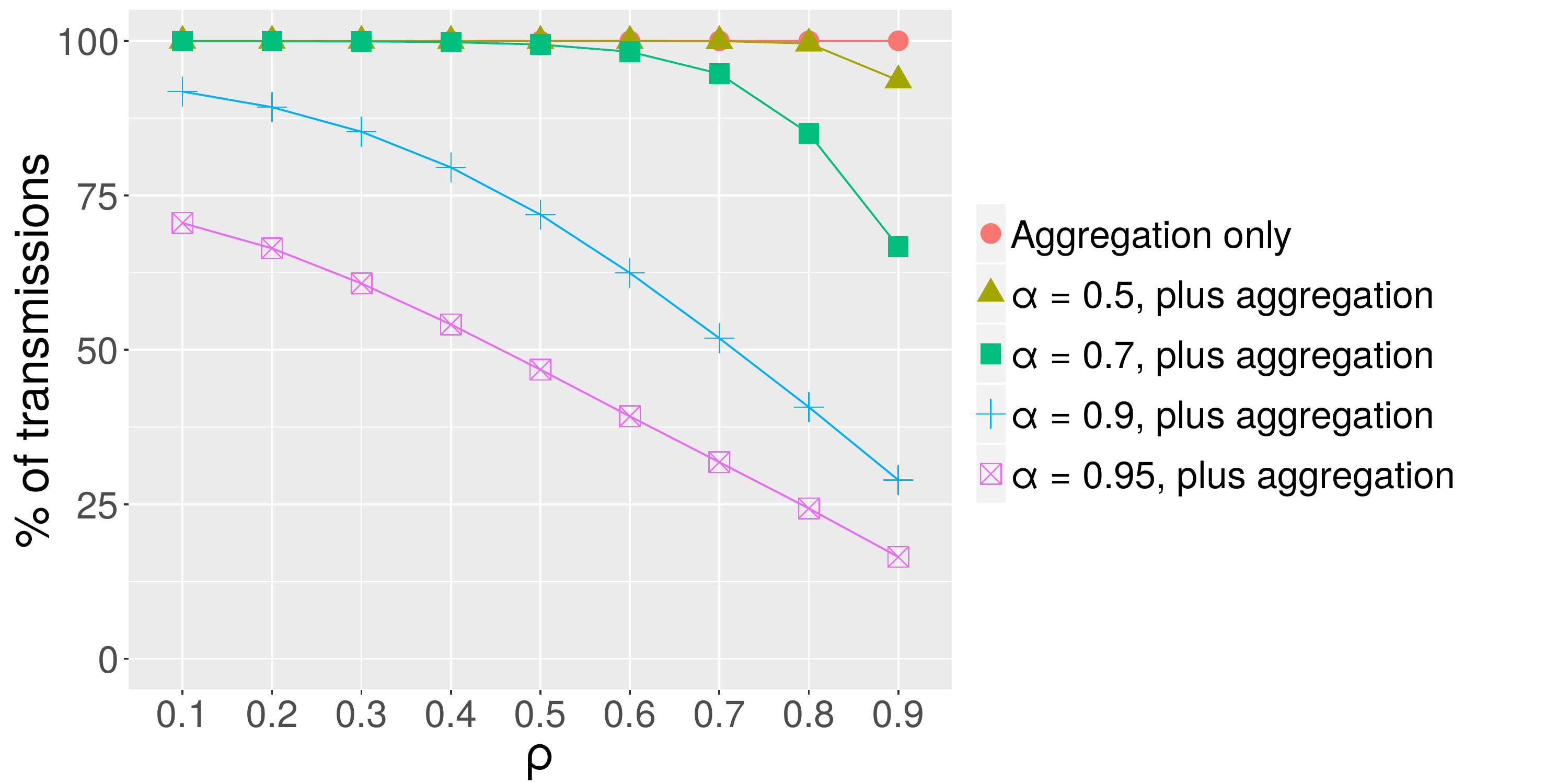}
	\caption{The effectiveness of the aggregations depend on the 
correlation between the measurements in a \emph{sub-tree}.}
	\label{fig:tx-c-3-d-3}
\end{figure}

To detail the power of the prediction and aggregation schemes, we considered a 
sensor network with five rings in which the aggregation scheme could reduce to 
$12.5\%$ the number of transmissions, similarly to the most accurate 
predictions.
Figure~\ref{fig:tx-c-3-d-3} highlights the gains obtained by adopting both 
schemes, where $100\%$ of transmissions represent the case where only data 
aggregation is adopted (i.e., $12.5\%$ of the transmissions in a \gls{WSN} with 
no optimization).
First, we can observe that the number of transmissions can be reduced to 
$15\%$ of its maximum in the best scenario, where the predictions are highly 
accurate, and the measurements in the \emph{sub-tree} are highly correlated.
This represents around $1.88\%$ of the total number of transmissions, 
given that it is $15\%$ of the $12.5\%$ of transmissions that the data 
aggregation would do.

Additionally, we did not observe any significant gains when the predictions 
were less accurate (around $0.5$) nor when the predictions were more accurate 
(around $0.7$), and the correlations were less than $0.7$.
Finally, with an average correlation ($0.5$), increasing the accuracy from 
$0.5$ 
to $0.9$ reduced by $30\%$ the number of transmissions.
Meanwhile, with an accuracy of $0.5$, increasing the correlation from $0.5$ to 
$0.9$ reduced only by $6.5\%$ the number of transmissions, which illustrates 
that the impact of making accurate predictions is much higher than having a high
correlation between the measurements. 

\subsection{Energy consumption}

Based on the number of transmissions and receptions, we can use the model 
presented before to estimate the total energy consumption of a sensor node $i$ 
in ring $d$ as
\begin{equation}
 E(\text{En}''_{i,d}) = \Ptx{''_{i,d}} \Etx{} + \Prx{''_{i,d}} \Erx{} + \Etop{} + 
\Emin{}\text{,}
 \label{eq:e-line}
\end{equation}
where \Etx{} and \Erx{} are the extra energy consumption to 
respectively transmit and receive one packet, \Emin{} is the minimum energy 
necessary to keep sensor nodes working without transmitting and receiving 
anything, and \Etop{} depends on where the prediction models are chosen. 
If prediction models are independently chosen, $\Etop{} =  0$;
if prediction models are chosen in the \gls{GW}, 
$\Etop{} =  \Erx{} + I_{d} \Etx{}$; and
if prediction models are chosen in sensor nodes, 
$\Etop{} = I_{d} \Erx{} + \Etx{}$.

% and it was the one considered in the plots.
To illustrate the applicability of this model, we estimated the energy 
consumption in a \gls{WSN} after three days of operation and compared with the 
results obtained in our simulations.
For this estimation, we considered a homogeneous sensor network with $D = 
5$ and $C = 3$ (i.e., $75$ sensor nodes plus the \gls{GW}).
To obtain the values of \Etx{}, \Erx{}, and \Emin{}, we simulated three TelosB 
motes in OMNET++ transmitting and receiving data without making any predictions.
After one simulated day, we calculated the average values for each parameter.

So far, we did neither distinguish delays nor packet lengths used in
aggregated transmissions and receptions from the case without aggregation.
In fact, in a real implementation, these transmissions could be done in the 
same 
packet types if we adopted simple aggregation functions, such as the maximum, 
minimum and the average of the measurements.
However, larger packets would mean higher energy consumption to transmit and 
receive, in comparison with the non-aggregated transmissions.
Therefore, to show the extensibility of our model, we used packets with eight 
times the payload of the normal packets in the aggregated transmissions.

\begin{figure}[t]
        \centering
	\includegraphics[width=0.98\textwidth]{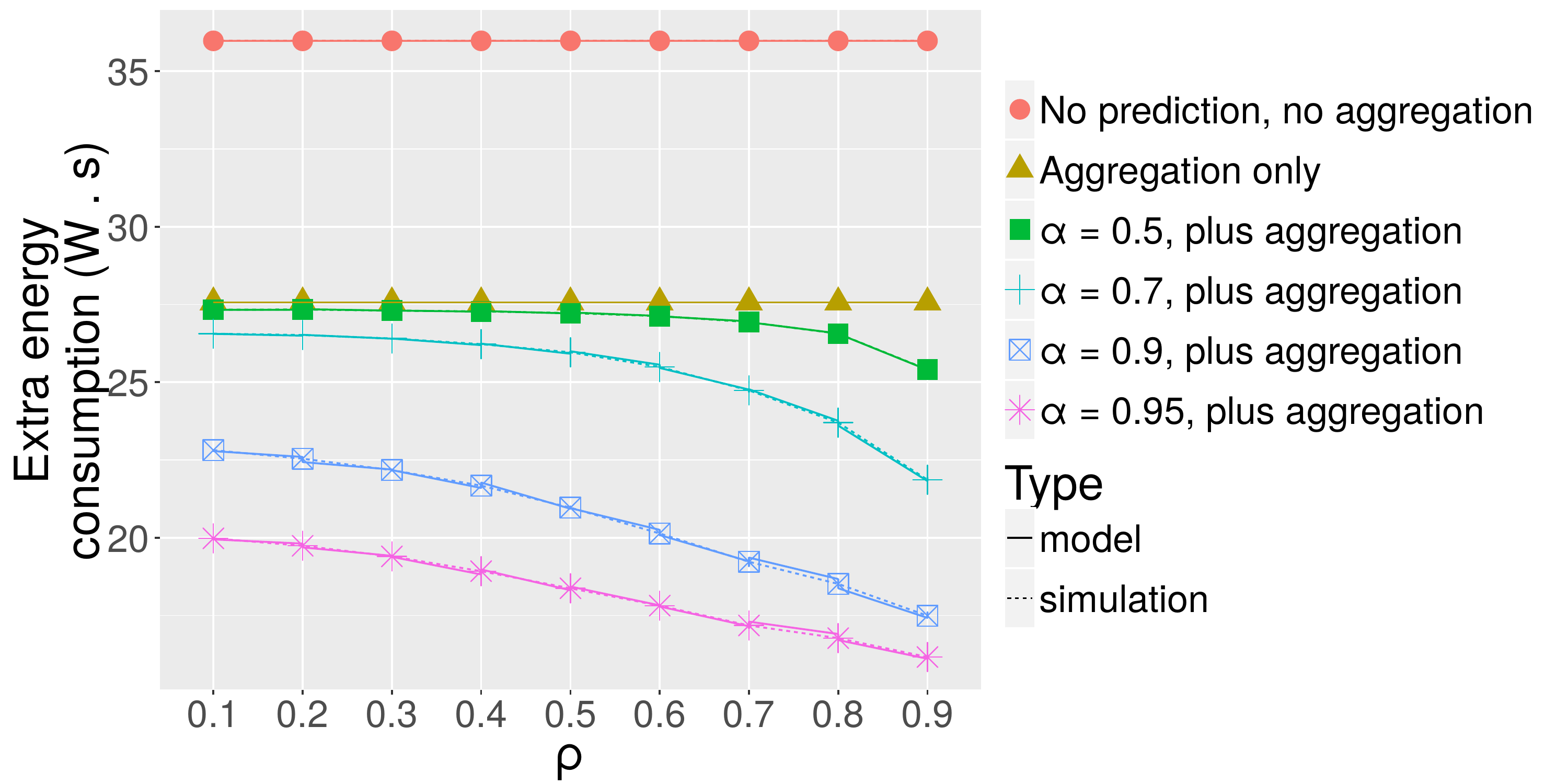}
	\caption{The model provides reliable results when compared with the 
simulations.}
	\label{fig:energy-comparison-c-3-d-3}
\end{figure}

To illustrate the results, we focused on the energy consumption of a 
sensor node in the first ring.
As sensor nodes in the first ring must handle the highest number of 
transmissions, they consume more energy than the others. 
As a consequence of such an energy consumption, these sensor nodes can run out 
of battery earlier than those in the other rings, which has a substantial
impact on the \glspl{WSN}' lifetime.
In Figure~\ref{fig:energy-comparison-c-3-d-3}, we can see that just adopting 
the aggregation scheme (without making predictions) reduces the extra 
energy consumption to $60\%$ of the total, yet larger packets are used.
The greatest gains, nonetheless, are obtained after adopting the \gls{DPS} 
and the aggregation scheme: they can save up to nearly $92\%$ of the energy 
consumed by the transmissions.
As explained before, the predictions' accuracy is more significant and 
has a higher impact than the correlation between the measurements in a 
\emph{sub-tree}.
Hence, a very low correlation ($0.1$) with highly accurate predictions ($0.95$) 
give better results than a high correlation ($0.9$) with an average accuracy 
($0.5$).

In fact, regardless of the values shown in the plot, the exact amount of 
saved energy depends on the hardware of the sensor nodes, their \gls{OS}, 
and the \gls{MAC} protocol in use, besides other configurations.
Nonetheless, the consumption is mainly driven by the relation between the 
minimum energy necessary to keep a sensor node making measurements and the 
amount of battery required for transmitting and receiving a packet.
In conclusion, the results presented here can facilitate the decision about 
adopting a \gls{DPS} in a \gls{WSN} with a similar arrangement, even if the 
sensor nodes' configurations differ from those considered in our investigation.

\section{Model validation}

In Section~\ref{sec:experimentation}, we designed an algorithm to calculate 
the impact of the data characteristics in the number of transmissions in 
scenarios where the \gls{WSN}'s structure follows the transmission 
model presented in Section~\ref{sec:proposed-mechanism}.
In this Section, however, our goal is to study a real use case, where the 
network is not as uniform as we assumed before, and the data is not exactly 
normally distributed but collected by real wireless sensor nodes in an 
experiment.

\subsection{Data selection}

\begin{figure}[t]
	\centering
	\includegraphics[width=0.8\textwidth]{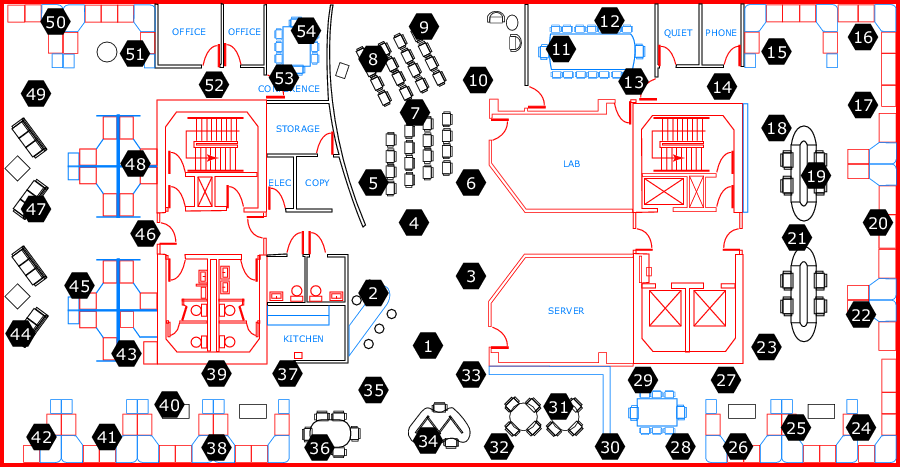}
	\caption{In the experiments made in the Intel Lab, temperature, relative 
	humidity, and luminance were measured by $54$ sensor nodes arranged
	as shown in this map~\cite{IntelLabData:2004:Misc}.}
	\label{fig:model-validation-data-selection}
\end{figure}

To validate the model, we adopted the data collected in the Intel 
Lab in the experiments described by Bodik et 
al.~\cite{IntelLabData:2004:Misc} and illustrated in 
Figure~\ref{fig:model-validation-data-selection}. 
In that experiment, $54$ sensor nodes transmitted measurements 
of temperature, relative humidity, and luminance every $30$ 
seconds during $37$ consecutive days.

Because there were several missing values in the original dataset,
we generated a new dataset by selecting a random measurement 
of each sensor in every interval of five minutes for eight days in a row.
We randomly selected nine sensor nodes that reported in more than $95\%$ 
of the time: $21$, $22$, $26$, $31$, $38$, $40$, 
$42$, $45$, and $46$.
In this new scenario, the network is not as uniform as we assumed before,
which challenges the applicability of our model.

From the new dataset, we selected eight consecutive days to calculate the data 
characteristics, i.e., the average and the standard deviation of the temperature 
measurements made by each sensor in each hour of each day.
%%% average correlation coefficient: fisherz2r(mean(fisherz(correlations)))
To estimate the average correlation coefficient between the measurements made 
by these sensor nodes, we first calculated the correlation matrix computed for all 
possible pairs of sensor nodes, applied the Fisher's z-transformation to transform 
these correlations into z-scores, calculated the average of the values 
and, finally, backtransformed the average to an r-score~\cite{Silver1987}.
For the temperature values, we calculated an average correlation coefficient of
$0.820068$.

Based on the average and the standard deviation values, we used the z-table 
to calculate four acceptance thresholds based on the expected accuracy 
levels: $0.5$, $0.7$, $0.9$, and $0.95$.
After calculating the average and four acceptance thresholds per sensor node per 
hour of each day, we considered that if a measurement differed from
the average by less than the acceptance threshold, the node would not
have transmitted during those days.

We highlight that our focus at this point is to compare the results obtained 
with the model and not how to generate prediction models.
In practice, it would be necessary a data study over the days preceding these 
experiments to design prediction models that would accurately 
predict the temperature measurements.
Therefore, due to a lack of space, we assume that those prediction models
could have been computed using state-of-the-art algorithms for predictions
and the data collected in the period before the experiments.
Given that, the \gls{GW} could disseminate these models, for example, some 
hours before the day that the predictions would refer.

Finally, we calculated the total number of transmissions made by a sensor node
that would be connected to the \gls{GW} and compared it with the case in 
which the sensor nodes would transmit every five minutes regardless the 
predictions.
Assuming a deployment where only one of the selected sensor nodes was directly 
connected to a \gls{GW}, every transmission would go through that sensor node,
and there would be necessary $20736$ transmissions to send all 
measurements to the \gls{GW} during the eight days considered in the
experiment.

\subsection{Results}

%% explain that the selected nodes could represent a branch in the
%% model presented before
Recall the \gls{WSN} model shown in Section~\ref{sec:proposed-mechanism},
where each branch is connected to the \gls{GW} via one sensor node.
It was possible to infer the number of transmissions at that sensor node
considering only the number of children it would have.
Such a value was defined as $K_{1}$ and depends on the number of
rings in the network.
For example, in a \gls{WSN} with $D$ rings,  $K_{1} = D^2 - 1$.
Therefore, given that we are working with $K_{1} = 8$ sensor nodes, 
a \gls{WSN} with three rings would have the same number of 
transmissions between the \gls{GW} and the sensor node which is closest 
to the \gls{GW}.
In other words, the nine sensor nodes selected for our observations could
represent a branch in the model presented before with $D = 3$ and 
$C \geq 3$.
Given that every branch in the model is expected to have the same number 
of sensor nodes and transmissions, the percentage of saved transmissions
will be also similar to the percentage of saved transmissions calculated
for a complete \gls{WSN}.

In an optimized deployment where a sensor node aggregates 
measurements from its children and transmits only once every five minutes, 
it would be necessary $2304$ transmissions to send all measurements to the 
\gls{GW} during the eight days considered in the
experiment.

Alternatively, if we adopted the \gls{DPS} besides the aggregation scheme,  
each transmission would be conditioned to the inaccuracy of (at least) one 
prediction in that transmission's five-minute window.
Table~\ref{table:model-validation-results-temperature} shows 
the number of temperature measurements transmitted 
using \gls{DPS} with aggregation, considering a deployment with
aggregation as a reference ($100\%$) for the maximum number of 
transmissions.

%\subsection{Results}

\begin{table}[t]
\centering
\def\arraystretch{1}
\rowcolors{2}{gray!25}{white}
\begin{tabular}{
>{\centering\arraybackslash}m{1.8cm}
>{\centering\arraybackslash}m{1.8cm}
>{\centering\arraybackslash}m{1.8cm}
>{\centering\arraybackslash}m{1.8cm}
}
{\cellcolor{gray!65}\bf Accuracy} & 
{\cellcolor{gray!50}\bf Real data} & 
{\cellcolor{gray!50}\bf Model} &
{\cellcolor{gray!50}\bf Difference} \\  
\hline 
%\cellcolor{gray!15}$50\%$ & $89.1\%$ & $89.2\%$ & $+0.1\%$ \\
%\cellcolor{gray!40}$60\%$ & $78.8\%$ & $77.3\%$ & $-1.5\%$ \\
%\cellcolor{gray!15}$70\%$ & $64.3\%$ & $62.5\%$ & $-1.8\%$ \\
%\cellcolor{gray!40}$80\%$ & $47.7\%$ & $45.5\%$ & $-2.2\%$ \\
%\cellcolor{gray!15}$90\%$ & $24.9\%$ & $26.0\%$ & $+1.1\%$ \\
%\cellcolor{gray!40}$95\%$ & $8.7\%$ & $14.5\%$ & $+5.8\%$ \\
\cellcolor{gray!15}$50\%$ & $87.3\%$ & $92\%$ & $4.6\%$ \\
\cellcolor{gray!40}$60\%$ & $78.8\%$ & $81.1\%$ & $2.3\%$ \\
\cellcolor{gray!15}$70\%$ & $67.4\%$ & $66.4\%$ & $-1.1\%$ \\
\cellcolor{gray!40}$80\%$ & $51.1\%$ & $48.7\%$ & $-2.4\%$ \\
\cellcolor{gray!15}$90\%$ & $27.6\%$ & $28\%$ & $0.5\%$ \\
\cellcolor{gray!40}$95\%$ & $11.2\%$ & $15.8\%$ & $4.5\%$ \\
\hline
\end{tabular}
\caption{Total number of transmissions considering different predictions' 
accuracy levels for the temperature measurements.}
\label{table:model-validation-results-temperature}
\end{table}

As expected, we could observe some difference between the model and
the number of transmissions expected in the real scenario.
We believe that this variation is mainly because the model is 
built based on the assumption the data is normally distributed, which 
was not true in the real experiment. 
This change impacts the final number of transmissions and partly 
explains the differences  between the number of transmissions obtained
using the data from the real experiment and the model.
Apart of that, the model seems still valid as 
a close approximation for this real case because the absolute 
difference between 
the results obtained using real data and those calculated using the
model were smaller than $2.5\%$ in most of the cases and smaller
than $5\%$ in the worst case.

\section{Related work}
\label{sec:related-work}

The mechanism called BBQ assumes that sensors nearby are likely to 
have correlated readings, which may mean that most of the measurements 
provide little benefit in approximate answers' quality~\cite{Deshpande2004}.
BBQ approximates the probability density function of the measurements to 
\gls{MVN} distributions and, given the correlation between known 
measurement(s) and the unknown one(s), it calculates their expected value 
associated with a confidence interval. 
If the confidence level is greater or equal than a user defined threshold, it 
answers the queries locally, without triggering new transmissions from sensor 
nodes.
In an outdoor scenario with less interference from humans and machines (where 
the sensed data approximates better to Normal distributions), it was possible
to achieve a significant reduction in the number of transmissions.
That improvement represented a reduction of $97\%$ in the energy consumption 
of the nodes and an 
acceptable level of mistakes (nearly $5\%$ of wrong answers).

Similarly to the model that we propose, the authors incorporated statistical 
models of real-world processes and exploited the correlation between 
measurements taken in the same vicinity.
They also incorporated the correlation between different types of data 
that the sensor nodes may be able to measure, for example, their voltage 
and the local temperature. 
On the other hand, the authors focused on modeling the number of transmissions 
in existing \glspl{WSN} to optimize their mechanism of query answering.
Our model is intended to model generic \glspl{WSN} used to report 
measurements periodically, based on their expected characteristics and takes 
into account the relative importance of nodes close to the \gls{GW}, i.e., 
those that must handle the highest number of transmissions and, therefore, are 
the bottleneck of \glspl{WSN}' workload.

Intanagonwiwat et al. have already modeled the impact of 
data aggregation schemes in \glspl{WSN}~\cite{Intanagonwiwat2001}.
More specifically, they showed the impact of network density on data 
aggregation 
using directed diffusion with a greedy algorithm to construct trees.
In their work, they proposed the in-network aggregation (along with the greedy tree) 
and compared the perfect aggregation (which saves the headers and merges the 
content in new packets with the same length) with the linear aggregation (which 
also saves the headers, but appends the content in larger payloads).
According to simulation results, the greedy algorithm can reduce $36\%$ of the 
energy consumption using linear aggregation and $43\%$ using the perfect 
aggregation in high-density networks.
Finally, the authors concluded that, in high-density networks, more energy is 
saved by the greedy algorithm, and the delays are as good as using the 
opportunistic algorithm.

More recently, Fan et al. focused on calculating the most energy efficient 
deployment strategy for \glspl{WSN} using an integral programming 
model~\cite{Fan2014}.
They used a regular hexagonal cell architecture~\cite{Li2006,Subir2009} to 
model the location of sensor nodes in the plane, which, in fact, is similar to a 
ring model with $C = 6$.
Using this model, they formulated the energy consumption of sensor nodes and 
\glspl{GW} based on the energy used to transmit, receive and process data in 
sensor nodes.
Finally, they also evaluated the impact of data aggregation in the energy 
consumption of the sensor nodes.

The works above inspired our model for data transmissions.
However, different from our proposal, they focused only on data aggregation 
and did not evaluate the effects of \glspl{DPS} in \glspl{WSN}.
% \glspl{DPS} started to be incorporated to \glspl{WSN} more recently.
% For example,
%%% The Implementation of an Adaptive Data Reduction Technique for Wireless
%%% An Application-Specific Forecasting Algorithm for Extending WSN Lifetime
% Notably, some works
Concerning the evaluation of \glspl{DPS}, Liu et al. were the first authors to 
introduce statistical methods to choose which prediction model better fits a 
certain environment~\cite{Liu2005}.
They created a formula to estimate the Prediction Cost ($\text{PC}$), which 
considers the percentage of transmitted measurements ($r$) and the user 
desired level of accuracy ($\alpha$)~\cite{Liu2005}.
More recently, an extended formula was designed and implemented in real sensor 
nodes to compare the savings using several prediction methods, such as the 
\Constant{}, \gls{ES} and \gls{ARIMA} 
methods~\cite{Aderohunmu2013Impl,Aderohunmu2013}.
% In their particular use case, the \Constant{} prediction method was the best 
% trade-off between accuracy and energy consumption in sensor 
% nodes.
The new formula is more generic than the original proposed and also considers 
the prediction models' memory footprint ($Ec$) as a significant computational 
cost for sensor nodes:

\begin{equation}
\text{PC} = [~\alpha f(e)~+~(1 - \alpha) r~]~\text{Ec,}
\end{equation}
where $e$ is the measure of the predictions' accuracy (e.g., \gls{MSE}, 
\gls{RMSE}, \gls{sMAPE}) and $f(e)$ is the accuracy according to the chosen 
measure.
The formula of PC can be useful to decide for adopting a \gls{DPS} or not, 
but it is limited to evaluate if predictions will save energy in one sensor 
node and does neither consider the impact of occasional aggregations, nor the 
work to forward transmissions from children nodes.

To the best of our knowledge, this is the first work that models the impact of
\glspl{DPS} in sensor networks and shows their potential as a whole class 
of applications.

\section{Conclusion and future work}
\label{sec:conclusion}

In this work, we presented a mathematical framework to calculate the gains 
and benefits of adopting a \gls{DPS} to reduce the number of transmissions in a 
\gls{WSN}.
% Using the proposed model, we showed that the benefits of adopting an 
% aggregation scheme are greater than using only predictions in larger \glspl{WSN} 
% and that combining both strategies leads to the highest savings.
% Moreover, we observed the most significant savings when we made accurate 
% predictions in the \gls{GW} and aggregated intermediate transmissions in the 
% sensor nodes. 
% Finally, our simulations also showed that the predictions' accuracy has 
% a higher impact than the measurements' correlation in the total number of 
% transmissions.
%
The model shows that, concerning the number of transmissions, the benefits of 
adopting an aggregation scheme are greater than using only predictions, and that 
combining both leads to the highest savings.
For example, as observed in Figure~\ref{fig:tx-c-3-d-3}, the number of 
transmissions and receptions in the bottlenecks (i.e., the sensor 
nodes in the first ring) can be reduced by nearly $98\%$ using accurate 
predictions and data aggregation.
Our simulations also showed that the accuracy of the predictions impacts 
more than the correlation between the measurements from different sensor 
nodes.
Finally, we compared the values obtained using our model with those that would
have been obtained using real data, and we observed that our model 
provides a good estimation of the number of transmissions that might happen 
in practice.

The main contribution for the future generations of \glspl{WSN} is a model that 
relies on the statistical theory to show the impact of sensor nodes' 
hardware evolution and the integration of \glspl{WSN} into the \gls{IoT}, which 
can be exploited to make predictions with higher accuracy in \glspl{DPS}.
% , the model provides means to endorse the adoption of 
% predictions to reduce the number of transmissions in sensor networks and 
% extend their lifetime.
The backbone of the model consists of an application of two statistical 
theorems:\begin{inparaenum}[(i)] 
 \item the Central Limit Theorem, which supports the normalization of the data 
measured by the 
sensor nodes; and
\item the Law of Large Numbers, which allows the extrapolation to similar 
scenarios based on the average number of neighbors and furthest distance from 
sensor nodes to \glspl{GW}.
\end{inparaenum}
% Thus, this model can be mainly used to exploit characteristics of \glspl{WSN} to 
% adopt predictions and improve the utilization of the channel resources.
% Additionally, the proposed model can be extended to calculate 
% the sensor nodes' energy consumption and estimate \glspl{WSN}' lifetime.

Considering that this model has been designed to represent different types of 
sensor networks, there are some challenges to set up the best parameters for 
each use case. 
From our experiments, we expect that the critical points in a real scenario may 
vary between:\begin{inparaenum}[(i)] 
 \item finding the precise correlation between measurements from several sensor 
nodes;
 \item approximating the measurements to Normal distributions, since it may 
require some data analysis in advance;
 \item having restrictions about changing the radio operation of some sensor 
nodes, because they may be fundamental to the sensor network connectivity, and 
the timeliness of the data delivered by the \gls{WSN}; and
 \item calculating the energy necessary for each step (i.e., transmission, 
reception, etc.), because details in the software and hardware implementations 
may influence such values~\cite{Pham2014}.
\end{inparaenum}

In future works, we plan to include this model in a larger system that can 
improve the accuracy of the predictions made by the \gls{GW} after consulting 
external sources of information.
With such an extended access to external information and using the existing 
formulas to estimate the prediction 
costs~\cite{Liu2005,Aderohunmu2013Impl,Aderohunmu2013}, \glspl{GW} will be 
able to choose the best mechanism to evaluate the quality of the measurements 
provided by their \glspl{WSN}.
Hence, all these features will make it possible to run self-managed systems 
that adapt the sensor network's operation according to their surroundings (and 
not only based on the observed environment) to achieve the best results, i.e., 
highest quality of measurements and fewest transmissions possible.

\section*{Acknowledgment}

This work has been partially supported by the Spanish Ministry of Economy and 
Competitiveness under the Maria de Maeztu Units of Excellence Programme 
(MDM-2015-0502), TEC2012-32354 (MINECO/FEDER) and by the ENTOMATIC
FP7-SME-2013 EC project (605073).

\section*{Bibliography}

\bibliography{bibliography}

\appendix

\section{Minimum average accuracy}
\label{appendix:accuracy}

Let us assume a monitoring \gls{WSN} with homogeneous sensor 
nodes periodically transmitting measurements every $1/f$ seconds, and a sensor 
node $i$ in ring $d$. 
According to (\ref{eq:x_d}), the number of transmissions at $i$ in a period of 
$1/f$ seconds is the sum $\Ptx{_{i}}~+~\Prx{_{i}}$.
If a \gls{DPS} is adopted, the number of transmissions will be 
$\Ptx{'_{i}}~+~\Prx{'_{i}}~+~\Xtop{}$, as defined in (\ref{eq:x-line}).
Therefore, a \gls{DPS} will reduce the number of transmissions during a period 
$T$ if and only if:

\begin{equation}
(\Ptx{'_{i}} + \Prx{'_{i}}) f T + \Xtop{} \leq  (\Ptx{_{i}} + \Prx{_{i}}) f T
\end{equation}

Thus, we can define $\alpha_{\text{min}}$ as the minimum average between the 
accuracies of the children of $i$ that would reduce the number of transmissions 
in a \gls{DPS} (without aggregation).
It must satisfy the following equation:

\begin{equation}
(\Ptx{'_{i}} + \Prx{'_{i}}) fT + \Xtop{} =  (\Ptx{_{i}} + \Prx{_{i}}) fT \\
\end{equation}

Based on (\ref{eq:ptx_d}), (\ref{eq:prx_d}), (\ref{eq:s-line-max}), and
(\ref{eq:r-line-max}), we can calculate it as:
\begin{equation}
 ((K_{d} + 1) ~\alpha^c_\text{min}  + (K_{d}~\alpha^c_\text{min}) 
) f T + \Xtop{} = ((K_{i,d} + 1)  + K_{d} ) f T
\end{equation}

Knowing this, we can work out the equation:
\begin{equation}
\begin{split}
%  \Ptx{'_{i}} + \Prx{'_{i}} + \Xtop{} &\leq  \Ptx{_{i}} + \Prx{_{i}} \\
((K_{d} + 1) ~\alpha^c_\text{min}  + (K_{d}~\alpha^c_\text{min}) 
) f T &= ((K_{d} + 1)  + K_{d} ) f T - \Xtop{} \\
\alpha^c_\text{min} (D^2 ~  + (D^2 - 1) ) f T &= (D^2  + 
(D^2 - 1) ) f T - \Xtop{} \\
 \alpha^c_\text{min} (D^2 ~  + (D^2 - 1) ) f  T &= (D^2  + 
(D^2 - 1) ) f T - \Xtop{} \\
 \alpha^c_\text{min} &= \frac{(D^2  + (D^2 - 1) ) f T - \Xtop{}}{
 (D^2 ~  + (D^2 - 1) ) f  T} \\
 \alpha^c_\text{min} &= 1 - \frac{\Xtop{}}{(D^2 ~  + (D^2 - 1) ) f T} \\
 \alpha_\text{min} &= \frac{\Xtop{}}{(D^2 ~  + (D^2 - 1) ) f T} \\
 \alpha_\text{min} &= \frac{\Xtop{}}{(2D^2 ~ - 1) f T} %\\
%  \alpha &> \frac{\Xtop{}}{(2D^2 ~ - 1) f T}
\end{split}
\end{equation}

Given that we assume no aggregation, the value of \Xtop{} is defined 
by~(\ref{eq:n-dis-gw-aggregation}):

\begin{equation}
\Xtop{} = \Ptx{^*} + \Prx{^*} = (2D^2 ~ - 1)\text{.}
\end{equation}
Therefore, 
\begin{equation}
\alpha_\text{min} = 1 / {f T}
\end{equation}

\section{Data model}
\label{appendix:data}

A Normal distribution is characterized by its probability density function 
whose pattern is often encountered in several types of observations.
According to the Central Limit Theorem, the sampling distribution of 
the mean of any independent random variable tends to be Normal, even if the 
distribution from which the average is computed is decidedly non-Normal.
For example, it has been shown that environmental readings--such as 
temperature, light, and humidity--done by outdoor WSNs can be approximated to 
normal distributions if properly managed~\cite{Deshpande2004}.

We will assume that a sensor network is composed of a set of sensor nodes $S$ 
and each sensor node $i \in S$ is responsible for measuring a 
certain parameter from the environment, such that the set of observations 
follows a Normal distribution with mean $\mu_i$ and variance $\sigma^2_i$. 
By convention, this is represented as $Y_i = N(\mu_i, \sigma^2_i)$.
A prediction $\bar{y}_i$ (for example, $\bar{y}_i = \mu_i$) can be calculated 
by the sensor node $i$ and the \gls{GW}.
We define the acceptance threshold $\varepsilon_i$, i.e., the prediction is 
told to be correct if the real observation ($y_i$) is in the interval 
$[\bar{y}_i~-~\varepsilon_i,~\bar{y}_i~+~\varepsilon_i]$.

Assuming that the data is normally distributed, the chances of observing a new 
value inside the accepted interval can be calculated by normalizing the value 
of $\varepsilon_i$, i.e., rewriting it in terms of the variance $\sigma_i^2$.
The normalized value of $\varepsilon_i$ is represented by $z_i$ as

\begin{equation}
z_i = \frac{\varepsilon_i - \bar{y}_i}{\sigma_i}
\label{eq:z_1d}
\end{equation}
Thus, in this case, the accuracy of the predictions ($\alpha_i$) can be 
calculated based on the cumulative distribution function of the normal 
distribution:

\begin{equation}
\Phi_{\mu, \sigma}(x) = {\frac {1}{2}}\left[1+\operatorname {erf} 
\left({\frac {x-\mu }{\sigma {\sqrt {2}}}}\right)\right]\text{.}
\end{equation}

Again, according to the Central Limit Theorem, we assume unbiased predictions 
and normally distributed errors.
Therefore, the percentage of observations that will fall outside the accepted 
interval is represented by the two-tailed Z-test (i.e., $2 \Phi(-|z_i|)$), and 
$\alpha_i$ is 
\begin{equation}
\alpha_i = 1 - 2 \Phi(-|z_i|)\text{.}
\label{eq:alpha_1d_one}
\end{equation}
By substituting the Equation~\ref{eq:z_1d} into Equation~\ref{eq:alpha_1d_one}, 
we can observe that

\begin{equation}
\alpha_i = 1 - 2 \Phi\left(-\left|\frac{\varepsilon_i - 
\bar{y}_i}{\sigma_i}\right|\right), 
\end{equation}
which shows that the accuracy of the predictions depends on the acceptance 
threshold, the mean and the variance of the data.

\section{Proof of $1 - \alpha^{x} \leq x~(1 - \alpha)$}
\label{appendix:inequality}

Let us assume two values $\alpha$ and $x$ such that $\alpha \in [0, 1]$ and $x 
\geq 1$. We want to show that $1 - \alpha^{x} \leq x~(1 - \alpha)$:

\begin{equation}
 \begin{split}
  1 - \alpha^{x} &\leq x~(1 - \alpha) \\
  1 - \alpha^{x} &\leq x - \alpha~x \\
  - \alpha^{x} + 1 &\leq x - \alpha~x \\
  \alpha^{x} &\geq 1 - x + \alpha~x \\
  \alpha^{x} &\geq 1 + x~(\alpha - 1)
 \end{split}
 \label{eq:inequality-methods}
\end{equation}

When $\alpha = 0$ or $\alpha = 1$, we can easily observe that the affirmation 
is true because of $x \geq 1$ by definition.
For the other values of $\alpha$, we can use the Bernoulli's 
inequality: %~\cite{Bernoulli1744}:
\begin{equation}
(1~+~i)^j \geq 1+ij\text{,}
\end{equation}
where $i > -1$, $i \neq 0$ is a real number and $j \geq 2$ an integer value.
Substituting the values of $\alpha$ and $x$ in 
Equation~\ref{eq:inequality-methods} respectively by $i + 1$ and $j$, the 
claim is proved.

\end{document}